\documentclass[12pt,a4paper]{article}

\pdfoutput=1
\usepackage[margin=2cm]{geometry}

\usepackage{amsmath,bbm}
\usepackage{amssymb}
\usepackage{graphicx} 
\graphicspath{{figures/}}
\usepackage[utf8]{inputenc}
\usepackage[T1]{fontenc}
\usepackage{mathrsfs}
\usepackage{mathtools}
\usepackage[dvipsnames]{xcolor}
\usepackage{xcolor}
\usepackage{hyperref}

\DeclareUnicodeCharacter{2212}{-}

\usepackage{tikz} 
\usetikzlibrary{chains,shapes,fit,calc}
\usetikzlibrary{positioning,arrows}

\usetikzlibrary{patterns,decorations.pathreplacing}
\usetikzlibrary{decorations.pathmorphing}
\usetikzlibrary{decorations.markings}
\usetikzlibrary{arrows,shapes}
\usetikzlibrary{shapes.misc}
\tikzset{
	gluon/.style={decorate, draw=black,
		decoration={coil,amplitude=4pt, segment length=4pt,aspect=0.7}} 
}
\tikzset{
	photon/.style={decorate, decoration={snake}},
}

\usepackage{float}
\usepackage{subfig}
\usepackage{floatpag}
\usepackage[T1]{fontenc}

\usepackage{upgreek}

    \makeatletter
    \let\@fnsymbol\@arabic
    \makeatother

\newcommand{\bea}{\begin{eqnarray}}
\newcommand{\eea}{\end{eqnarray}}
\newcommand{\be}{\begin{equation}}
\newcommand{\ee}{\end{equation}}
\newcommand{\ba}{\begin{array}}
\newcommand{\ea}{\end{array}}

\def\met{\ifmmode E_{T}^{miss} \else$E_{T}^{miss}$ \fi} 
\def\Zp{\ifmmode Z' \else $Z'$ \fi}

\def\gsim{\raise0.3ex\hbox{$\;>$\kern-0.75em\raise-1.1ex\hbox{$\sim\;$}}}
\def\lsim{\raise0.3ex\hbox{$\;<$\kern-0.75em\raise-1.1ex\hbox{$\sim\;$}}}

\begin{document}

\thispagestyle{empty}
\begin{flushright}
DESY 21-146 \\
ICAS 066/21 \\
IFIC/21-33
\end{flushright}
\vspace{0.1in}
\begin{center}
{\Large \bf 
$Z'$-explorer 2.0:  reconnoitering the dark matter landscape} \\
\vspace{0.2in}
	{\bf V\'ictor Mart\'in Lozano$^{(a)\dagger}$,
	Rosa Mar\'ia Sand\'a Seoane$^{(b)\star}$,
	Jose Zurita$^{(c),\diamond}$
}
\vspace{0.2in} \\
{\sl $^{(a)}$
Deutsches Elektronen-Synchrotron DESY, Notkestr. 85, 22607 Hamburg, Germany}
\\[1ex]
	{\sl $^{(b)}$ International Center for Advanced Studies (ICAS)\\
 UNSAM, Campus Miguelete, 25 de Mayo y Francia, (1650) Buenos Aires, Argentina }
\\[1ex]
{\sl $^{(c)}$
Instituto de F\'{\i}sica Corpuscular, CSIC-Universitat de Val\`encia, E-46980 Paterna, Valencia, Spain}
\end{center}
\vspace{0.1in}

\begin{abstract}
We introduce version 2.0 of  $Z'$-explorer, a software tool which provides a simple, fast and user-friendly test of models with an extra $U(1)$ gauge boson ($Z'$) against experimental LHC results. The main novelty of the second version is the inclusion of missing energy searches, as the first version only included final states into SM particles. Hence $Z'$-explorer 2.0 is able to test dark matter models where the $Z'$ acts as an s-channel mediator between the Standard Model and the dark sector, a widespread benchmark employed by the ATLAS and CMS experimental collaborations. To this end, we perform here the first public reinterpretation of the most recent ATLAS mono-jet search with 139 fb$^{-1}$. In addition, the corresponding searches in the visible final states have also been updated. We illustrate the power of our code by re-obtaining public plots, and also showing novel results. In particular, we study the cases where the $Z'$ couples strongly to top quarks (top-philic), where dark matter couples with a mixture of vector and axial-vector couplings, and also perform a scan in the parameter space of a string inspired Stückelberg model. $Z'$-explorer 2.0 is publicly available on GitHub. 
\end{abstract}

\vspace*{2mm}
\noindent {\footnotesize E-mail:
{\tt 
$\dagger$ victor.lozano@desy.de,
$\star$ rsanda@unsam.edu.ar,
$\diamond$ jzurita@ific.uv.es.
}
}

\newpage

\section{Introduction}
\label{sec:intro}
Modern model-building requires confronting the parameter space (masses, couplings) of a New Physics model against a large and ever-growing variety of experimental constraints. The reinterpretation of LHC searches can then range from trivial to highly-involved to (almost) impossible. The difficulty of the task can depend on several factors, including the implicit and explicit assumptions of a given study, the degree of model-independence and the public availability of the necessary information to reproduce the reported bounds outside of the specific collaboration (e.g. experimental efficiencies). A fruitful dialogue between the phenomenological and experimental communities is already ongoing~\cite{Abdallah:2020pec} which has considerable helped the public reinterpretation effort.

We can classify the re-interpretation codes in two broad categories. First and foremost, the \emph{general} codes provide a framework where any LHC study can be implemented, and those results applied to arbitrary models, e.g. CheckMATE~\cite{Drees:2013wra,Dercks:2016npn}, MadAnalysis5~\cite{Dumont:2014tja,Conte:2018vmg} and GAMBIT~\cite{GAMBIT:2017yxo}. This flexibility on the model-independence comes with the price of requiring the generation of  a large number  MonteCarlo events in order to confront a single point in parameter space with the experimental data. The second category, \emph{dedicated} codes, narrows the applicability by focusing on a specific model framework (e.g.  HiggsBounds~\cite{Bahl:2021yhk} and HiggsSignals~\cite{Bechtle:2020uwn} for extended scalar sectors, or SModelS~\cite{Kraml:2013mwa} for simplified models), which allows to bypass the CPU-time consuming event generation and thus allow a fast exploration of the parameter space. In the latter category belongs our tool, $Z'$-explorer~\cite{Alvarez:2020yim} which focuses on models where the SM is augmented with a new $U(1)$ gauge boson, dubbed $Z'$.

 $Z'$-models are widely used in a variety of contexts, see e.g.~\cite{Leike:1998wr,Langacker:2008yv} for a review. Of our particular interest is a $Z'$ boson acting as a s-channel mediator to the dark sector~\cite{Buchmueller:2014yoa,Abercrombie:2015wmb}. The existence of the mono-X process $p p \to Z' X \to \chi \chi X$, where $X=j, \gamma, Z, W, \dots$ is a Standard Model (SM) particle that must be present for the event to be recorded on tape (``triggered''), requires that the $Z'$ couples to both the Standard Model and the dark matter candidate $\chi$. This provides a robust foundation to the existing mediator search programme at the LHC, encompassing both \emph{visible} (i.e. Standard Model) and \emph{invisible} (dark sector) final states. Having a fast and flexible tool for the simultaneous reinterpretation of the broad palette of experimental searches is a desirable addition to the model-builder's toolkit. A first step in that direction was done in $Z'$-explorer 1.0, where the whole suite of visible final states was implemented~\cite{Alvarez:2020yim}, requiring as user input only the $Z'$ mass and its coupling to SM particles, which avoids the need for event generation.

In this article we extend the capabilities of $Z'$ explorer to further include searches with missing energy. To that effect, we perform (to the best of our knowledge) the first public reinterpretation and validation of the ATLAS mono-jet study with 139 fb$^{-1}$ of total integrated luminosity~\cite{ATLAS:2021kxv}, which provides the most stringent constraints from the whole set of mono-X searches when $X$ comes from initial state radiation~\cite{Bernreuther:2018nat}. When compared to ZPEED~\cite{Kahlhoefer:2019vhz} (which included all $Z'$ decays into di-jet and di-lepton channels), $Z'$-explorer additionally includes the $WW$ and $Zh$ channels (from version 1.0) and the missing energy studies described in this article (version 2.0).

Our code goes beyond the public results presented by ATLAS on two directions. First and foremost, we allow for arbitrary couplings to every SM fermion. The oversimplifying assumption of having one single common quark coupling $g_q$ and/or a single lepton coupling $g_l$ can be dropped, revealing an interesting interplay between the visible and invisible sectors. We note that in this case, since $Z'$-explorer does yet not include constraints from flavor physics (which can nonetheless be easily obtained from e.g. {\tt flavio}~\cite{Straub:2018kue}), the user must taken them separately into account. Second, we consider a general coupling structure for the  $Z'-\chi-\chi$ vertex, going beyond the two benchmark setups commonly used by the experimental collaboration: the vector and axial-vector scenarios. Our results are implemented in $Z'$-explorer 2.0, which is publicly available on GitHub~\cite{github}. Since $Z'$-explorer does not require simulating events, it is appropriate for a thorough scanning the parameter space when compared to general re-interpretation codes. 

The current article is structured as follows. In section~\ref{sec:model} we review the fundamentals of $Z'$ models, presenting the parametrization used within $Z'$ explorer, which coincides with the one adopted by the Dark Matter Working Group (DMWG)~\cite{Abercrombie:2015wmb}.\footnote{Except that the DMWG employs the vector (V) - axial-vector (A) basis for the couplings, instead of the left-right used in $Z'$-explorer.} In section~\ref{sec:ATLAS} we present the validation of the ATLAS mono-jet study~\cite{ATLAS:2021kxv}. We exemplify the impact of $Z'$-explorer by applying it to a series of examples in section~\ref{sec:resu}, and we reserve section~\ref{sec:conclu} for our conclusions. Technical details regarding the software implementation are left for Appendix~\ref{app:code}.


\section{$Z'$ models: theoretical framework}
\label{sec:model}
In this section we present a brief summary of the theoretical framework (for details we refer the reader to~\cite{Buchmueller:2014yoa})   and describe in a nutshell the main features of the dark sector implementation in $Z'$-explorer 2.0, while the technical details are presented in Appendix~\ref{app:code}.

We augment the SM with a new $U(1)$ gauge boson $Z'$ and a SM singlet Dirac fermion $\chi$. The latter is our dark matter candidate, rendered stable by a discrete $\mathbf{Z}_2$ symmetry, as customary in dark matter models. The SM fermions $f$ and the dark matter $\chi$ have arbitrary left- and right-handed couplings, $g_{f,\chi}^{L,R}$ to the $Z'$ mediator. The relevant Lagrangian then reads
\be
\mathcal{L} \supset Z'_{\mu} \left[\sum_{f}\left( g_{f_L} \bar f_L \gamma ^{\mu} f_L + g_{f_R} \bar f_R \gamma ^{\mu} f_R \right) + g_{\chi_L} \bar \chi_L \gamma ^{\mu} \chi_L + g_{\chi_R} \bar \chi_R \gamma ^{\mu} \chi_R \right] \, ,
\label{eq:Zplag}
\ee
where the index $f$ runs over all the SM fermions (for the treatment of neutrinos see Appendix~\ref{app:code}). This Lagrangian coincides with that adopted by the ATLAS and CMS collaborations~\cite{Buchmueller:2014yoa} when all the leptonic couplings are set to zero, and when the $Z'$ has either  vector ($g_{f_L} = g_{f_R}$) or axial-vector  ($g_{f_L} =  - g_{f_R}$) couplings to the SM quarks and to the dark matter $\chi$. While it has been shown that  the different couplings in equation~\ref{eq:Zplag} must obey non-trivial constraints among them if unitarity and gauge invariance are imposed~\cite{Kahlhoefer:2015bea,Cui:2017juz}, we consider this framework a minimal parametrization of richer models, that could feature more than one mediator and/or new dark states beyond $Z'$ and $\chi$. Hence each left- and right-handed (or axial and vector) coupling is treated as a free parameter by $Z'$-explorer. We also contemplate the possibility of $Z'$ decays into new, non-SM states (other than $\chi \chi$) parametrized by an unknown decay width $\Gamma_{xx}$. 
In comparison with $Z'$-explorer 1.0, the new parameters the user must input are the dark matter mass $m_{\chi}$ and the dark matter couplings to the $Z'$, $g_{\chi_{L,R}}$.

A foundational ingredient of $Z'$-explorer's philosophy is to employ the \emph{expected} limits instead of the observed ones. This is an approximation that holds as long as no large fluctuations in the data are present. The rationale behind it is to be able to disentangle the experimental sensitivity of a given channel from these fluctuations. In its current form, $Z'$-explorer would not be appropriate in a discovery (or large significant fluctuation) scenario. We emphasize, again, that the goal of $Z'$-explorer is to set exclusion bounds on the parameter space given by equation~\ref{eq:Zplag}. Hence throught this work all limits and background events reported correspond to the \emph{expected} case, unless noted otherwise.

A key constituent of $Z'$-explorer is the narrow-width approximation (NWA), which allows to factorize the LHC processes involving $Z'$ as a \emph{production cross section} times a corresponding branching ratio in a given channel. It is then important to check if the approximation is fulfilled. The calculation of the $Z'$ partial width at leading order (LO)  is straightforward and full expressions are presented in Ref.~\cite{Buchmueller:2014yoa}. In the limit where $m_{f, \chi} \lesssim M_{Z'}$ the $Z'$ width is independent of the left- and right-handed coupling structure, reading
\be
\frac{\Gamma_{Z'} }{M_{Z'}} = \frac{1}{12 \pi} \left( g_{\chi}^2 + \sum_{f} N_C g_f^2 \right) \, ,
\label{eq:zpwidth}
\ee
where $N_C$ is the number of color of each SM fermion $f$: 3 for quarks and 1 for charged leptons. $Z'$-explorer does not stop the execution of the code, but prints a warning in the output if the width-to-mass ratio of equation~\ref{eq:zpwidth} is larger than 5 \%. For a leptophobic $Z'$ with a single $g_q$ coupling being equal to (4 times smaller than) $g_{\chi}$, this 5 \% threshold is reached for $g_q = 0.35 (0.25)$, which sets a ballpark value for these couplings. We note that even if the width is small, interference effects can be important if the lepton decay rates are appreciable, as discussed in reference~\cite{Kahlhoefer:2019vhz}. This work introduced the public code ZPEED. In comparison with $Z'$-explorer, ZPEED outputs the likelihoods for a given point in parameter space while including the relevant interference effects, but it does not include SM final states with gauge bosons and scalars ($W^+ W^-, Z h$) and neither does include invisible decays.

The model described by equation~\ref{eq:Zplag} has been implemented in the Universal Feynman Output~\cite{Degrande:2011ua} (UFO) format in reference~\cite{ufomodel}. Along this work, we will employ  {\tt MadGraph5\_aMC@NLO} \cite{Alwall:2014hca} for event generation at the parton level, {\tt Pythia 8} \cite{Sjostrand:2014zea,Sjostrand:2006za} for showering and hadronization, and {\tt Delphes 3.4.2} \cite{deFavereau:2013fsa} for detector simulation, using the default ATLAS card. We will work at leading-order (LO) accuracy in the strong coupling, which is justified since next-to-leading order QCD effects have been found to be quite mild~\cite{Backovic:2015soa}.

A series of sanity checks has been performed with the UFO model. On one hand, we have verified the validity of the partial widths expressions by numerically comparing them to those obtained with MadWidth~\cite{Alwall:2014bza}. On the other hand, we have studied several kinematical distributions (in particular missing transverse energy, and the leading jet momenta and the jet pseudorapidities) to verify that, as expected, the shape of these distributions (which are those employed by ATLAS and CMS) depends only upon $M_{Z'}$ and $m_{\chi}$, while being independent of all couplings. This is a crucial assumption upon which $Z'$-explorer relies. 

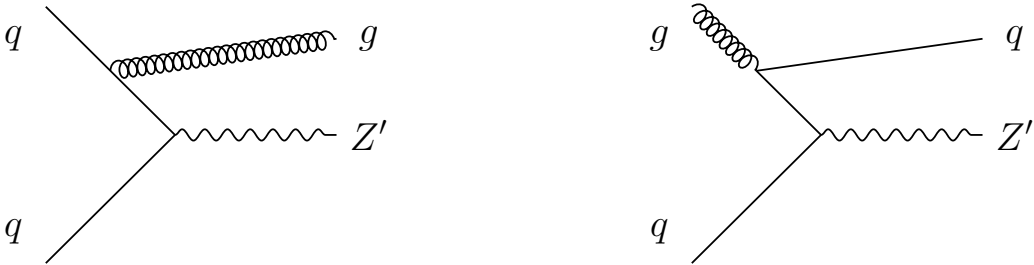
\begin{figure}[h]
	\begin{center}
		\scalebox{0.85}{
			\begin{tikzpicture}
			\begin{scope}[thick] 
			
			\draw[-] (0,2)--(2,0);
			\draw[-] (0,-2)--(2,0);
			\draw[photon] (2,0) -- (4.5,0);
			\draw[gluon] (1,1) -- (4.5,1.5);
			\node[black] at (5.0,0.00) {{\Large $Z'$}};
			\node[black] at (5.0,1.5) {{\Large $g$}};
			\node[black] at (-0.5,1.5) {{\Large $q$}};
			\node[black] at (-0.5,-1.5) {{\Large $q$}};

			\draw[-] (11,1)--(12,0);
			\draw[gluon] (10,2)--(11,1);
			\draw[-] (11,1)--(14.5,1.5);
			\draw[-] (10,-2)--(12,0);
			\draw[photon] (12,0) -- (14.5,0);
			\node[black] at (15.0,0.00) {{\Large $Z'$}};
			\node[black] at (9.5,1.5) {{\Large $g$}};
			\node[black] at (9.5,-1.5) {{\Large $q$}};
			\node[black] at (15.0,1.5) {{\Large $q$}};
			\node[white] at (15.0,2.50) {{}};
			\node[white] at (15.0,-2.250) {{}};
			\end{scope}
			\end{tikzpicture}
		}
	\end{center}
	\caption{Feynman diagrams of the $Z'$ production with a radiated jet from the inital state. This process could be initated either by a pair of quarks radiating a gluon (left) or a pair gluon quark where a quark is radiated (right).}
	\label{fig:ISRFD}
\end{figure}

The quark couplings obviously enter in the production cross section for the $p p \to Z' j$ process. The two main processes for $p p \to Z' j$ are shown in Fig.~\ref{fig:ISRFD}, in the first one a pair of quarks annihilate to produce a $Z'$ while a gluon is radiated from one of them (left diagram from Fig.~\ref{fig:ISRFD}). The other process is initiated by a gluon and a quark. The gluon splits into two quarks, one of them annihilates with the other inital quark to give a $Z'$ while the other is radiated. We have also verified that the different quark contributions from the initial state do not interfere with each other, namely that
\be
\sigma(p p \to Z' j) = \sum_{q_i} g_{q_i}^2 \left[ \sigma (q_i \bar{q_i} \to Z' g) +  \sigma(q_i g \to Z' q_i) +  \sigma(\bar{q}_i g \to Z' \bar{q}_i) \right] \, .
\label{eq:xs}
\ee
holds. We note that the different cross sections $\sigma$ in the square brackets are \emph{not} parton level matrix-elements, but they also include the corresponding convolution with the parton distribution functions (PDF). Here we assume that CP is conserved, hence the different between $q_i g$ and $\bar{q} g$ initial states is only due to the PDFs. 

For the fast evaluation of these production cross sections, we follow a procedure analogous to the one used in $Z'$-explorer 1.0. For the visible channels,  the relevant process under consideration is $q_i \bar{q_i} \to Z'$. Due to the trivial scaling with the $g_q$ couplings and the lack of interference among the different channels, it is enough to make a fine scan of $\sigma (q_i \bar{q_i} \to Z')$ in the 1-dimensional parameter space given by $M_{Z'}$. Here the task is slightly more complicated since now the different cross sections from equation~\ref{eq:xs} must be scanned in the two dimensional mass plane $M_{Z'}-m_{\chi}$. Nonetheless, this is not an obstacle, and with a fine scan of this mass plane the cross section of any given point can be numerically estimated with great accuracy, as described in Appendix~\ref{app:code}.


\section{ATLAS search for $Z'$ mediators to the dark sector}
\label{sec:ATLAS}
In this section we perform the validation of the ATLAS mono-jet search with 139 fb$^{-1}$ and explain the implementation in $Z'$-explorer 2.0. We comment on the shortcomings of our validation while making suggestions for the improvement of the reinterpretation material.

There is a large number of mono-X searches conducted by the ATLAS ~\cite{ATLAS:2021kxv,ATLAS:2021shl,ATLAS:2018nda,ATLAS:2017nyv,ATLAS:2020uiq}
and CMS collaborations~\cite{CMS:2021far,CMS:2017zts,CMS:2017jdm,CMS:2017qyo}. In our $Z'$ setup, with only one mediator and one dark matter particle, the $X$ particle comes exclusively from initial state radiation (ISR) and it has been shown that in that case the mono-jet search trumps over all the others~\cite{Bernreuther:2018nat}. We will then focus on the latest ATLAS mono-jet analysis~\cite{ATLAS:2021kxv}.\footnote{While this work was in the final stages of completion, updated monojet results by the CMS collaboration were made public~\cite{CMS:2021far}. Their sensitivity to the axial-vector scenario is comparable with the ATLAS result.}.  Nonetheless, for a successful validation of this channel we will also consider an older version of the study, using 36.1 fb$^{-1}$~\cite{Aaboud:2017phn} of total integrated luminosity, since each search present different benchmark points. The latter study had been implemented in the Physics Analysis Database (PAD) of MadAnalysis5~\cite{DVN/DFQPGU_2021}, which provided a useful cross-check of our signal cutflows~\footnote{We are indebted to Benjamin Fuks for providing us with MadAnalysis5 v.1.9.35.}. Nonetheless, due to the modular nature of $Z'$-explorer, and the goal to keep it as flexible as possible for future updates, we plan to implement the neglected mono-X studies in a future release including more general dark matter sectors. In that case,  final state radiation (FSR) also plays an important role and hence there is no ``a-priori''  predominance of the mono-jet channel.

The ATLAS study with 139 fb$^{-1}$ considers events that passes the $\met > 200$ GeV trigger, and requires the presence of at least one energetic jet $j_1$ fulfiling $p_{T,j_{1}}>250$ GeV and $|\eta|<2.4$, while up to a maximum of 4 jets with $p_T > 30$ GeV and $|\eta|<2.8$ are allowed. In all cases, there is a minimum azimuthal distance requirement between each jet and the direction of the transverse missing momentum, $\Delta\phi(jet,\vec{p_{T}^{miss}})>0.4 (0.6)$ for \met > 250 GeV ($\met \in [200-250]$ GeV). To complete the event selection, events with leptons ($e, \mu, \tau$) or photons are vetoed.

After this event selection, the analysis is perfomed using two sets of signal regions: an \emph{inclusive} selection (IM1, IM2, \dots) and an \emph{exclusive} one (EM1, EM2, \dots). As their names suggest, the inclusive analysis requires $\met$ above a certain threshold value, while the exclusive analysis requires an specific interval. The exclusive signal regions (which are those employed by $Z'$-explorer), with their $\met$ thresholds, are detailed in Table~\ref{tab:EM}. The previous version of this study uses a similar cutflow, albeit with fewer signal regions, tighter cuts on $\met$ and looser requirements on $e, \mu$ and no $\tau$ neither photon vetoes. Since the exclusion curves are only derived for the exclusive bins, 
we will restrict ourselves to the EMi regions. For completeness, we also present in Table~\ref{tab:EM} the expected (predicted) number of background events in each of them.

\begin{table}[h]
\centering
	\begin{tabular}{cccccccc}
	\hline
	Exclusive &  EM0 &  EM1 & EM2 & EM3  & EM4 & EM5 & EM6 \\ 
	$E_{T}^{miss}$ [GeV] & 200-250 & 250-300 & 300-350 & 350-400  & 400-500 & 500-600 & 600-700 \\ 
	Predicted & 1783000 & 753000 & 314000  & 140100  & 101600 & 29200  & 10000 \\ 	     
	\hline               
	Exclusive & EM7 & EM8 & EM9 & EM10 &  EM11 &  EM12 & \\
	$E_{T}^{miss}$ [GeV]  & 700-800 & 800-900 & 900-1000 & 1000-1100 & 1100-1200  & >1200  &\\
	Predicted  & 3870       &1640         & 754            & 359             & 182  &   \\
	\hline
	\end{tabular}
	\caption{Exclusive (EM0-EM12) signal regions defined in ATLAS monojet search at $13$ TeV and $139$ fb$^{-1}$, together with the expected number of background events.}
	\label{tab:EM}
\end{table}

For our validation, we turned to the public available material. The latest study uses as a benchmark point $M_{Z'} = 2000$ GeV and $m_{\chi} =1$ GeV, while the previous version employs $M_{Z'} = 1000$ GeV and $m_{\chi} =400$ GeV, and a universal coupling to quarks $g_q$ is set to 0.25, while $g_{\chi}$ is fixed to 1. These two benchmark points represent different kinematic regimes (in the former we can neglect the effect of $m_{\chi}$ while in the other the dark decay is close to the threshold), and hence the combination of both provide crucial information for the proper reinterpretation.

The ATLAS study only present results for the vector mediator and axial-vector mediator cases. ATLAS provides an official cutflow for one benchmark point (in the 139 fb$^{-1}$ version), and the $\met$ distribution for the signal benchmark and the background. The previous version of this study presented in addition (for the other benchmark point) the leading jet transverse momentum $p_T(j_1)$ distribution, which provides a welcome additional cross-check of the validation procedure.

Regarding the 95 \% confidence level (CL) exclusions, ATLAS presents the results for the observed upper limits (UL) on the product of the signal cross-section, acceptance and efficiency using the inclusive selection, and the  95 \% CL exclusion curves (``Brazilian flags'') using the exclusive selection. This creates an additional difficulty for the validation, as i) only the \emph{observed} UL are reported, and not the expected ones, ii) the use of different selections between the UL and the exclusion curves prevents the derivation of the Brazilian flags from the UL  and iii) correlations among the different bins are not reported.

\begin{figure}[h!]
 \centering
 \subfloat[]{\includegraphics[width=0.48\textwidth]{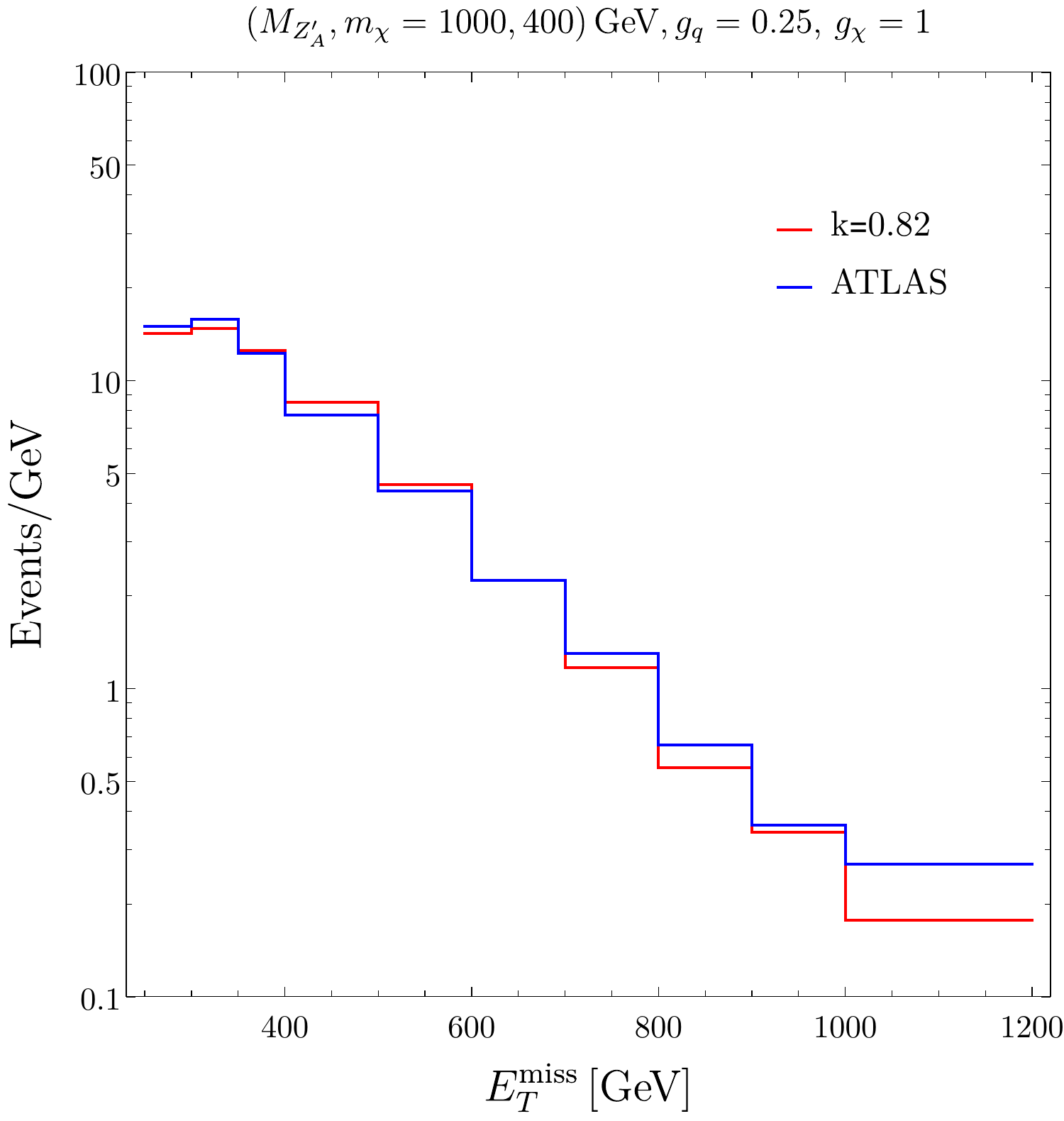}} \hspace{3mm}
 \subfloat[]{\includegraphics[width=0.48\textwidth]{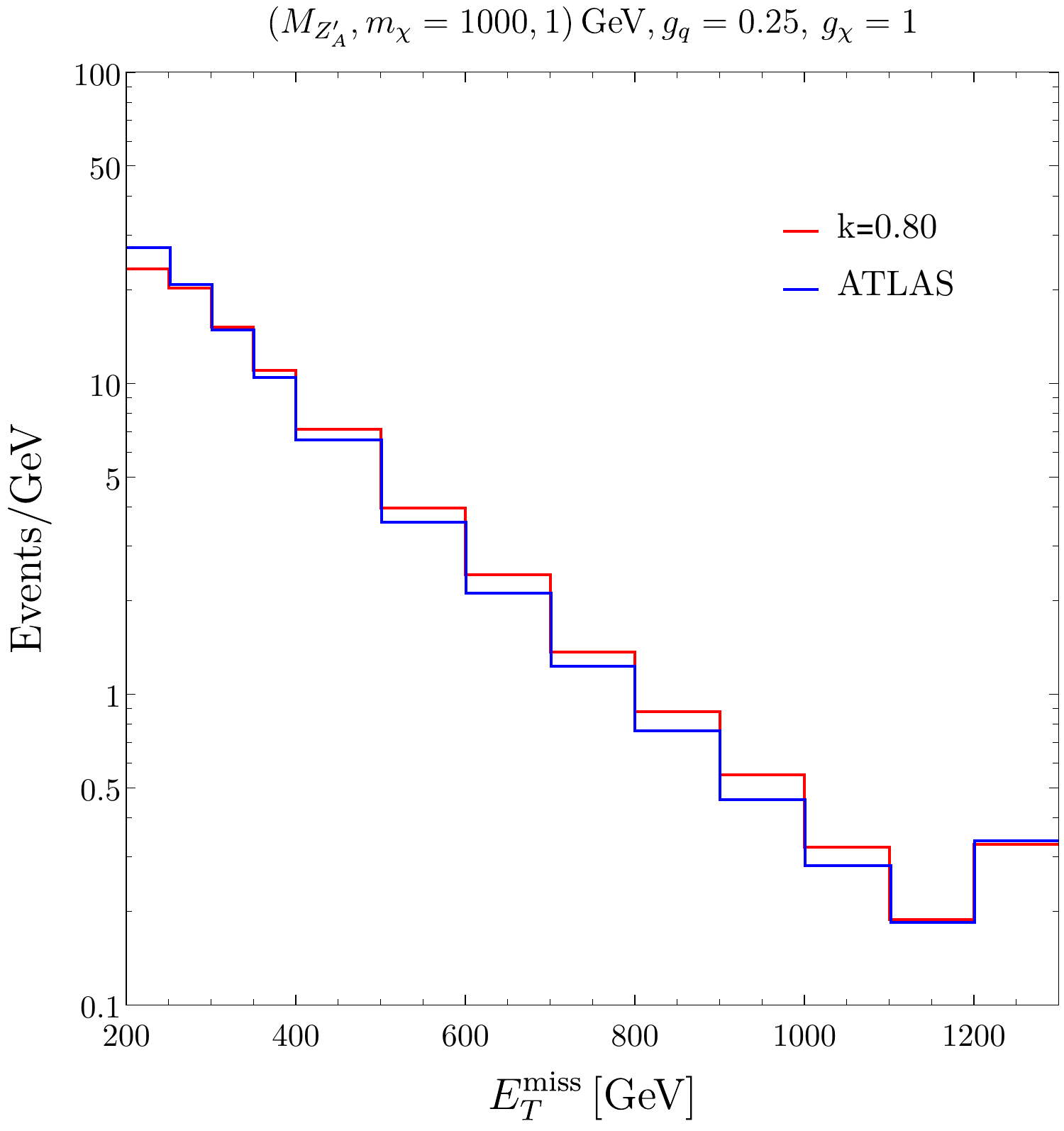}}
 \caption{\small Missing transverse energy distributions for a signal point with  $g_q=0.25$, $g_\chi=1$, and (a) $M_{Z'}=1000$ GeV, $m_{\chi}=400$ GeV; (b) $M_{Z'}=2000$ GeV, $m_{\chi}=1$ GeV in the axial-vector mediator scenario. A flat k-factor of 0.82 (0.80) has been applied in the left (right) panel.}
\label{fig:met_distros}
\end{figure}

We present in figure~\ref{fig:met_distros} the missing energy distributions obtained for the axial-vector case by ATLAS, together with the result of our simulations. In the left panel we present the 36 fb$^{-1}$ study, while in the right panel we present the results for 139 fb$^{-1}$. In order to match the published distributions, we need to include a global k-factor of $0.80 \pm 0.02$, whereas the older version (for a different benchmark point and with less background events, hence a larger statistical error) required $0.82 \pm 0.02$. We can then conclude that our event generation pipeline is validated, although as mentioned additional distributions, and potentially additional benchmark points would be a desirable addition, following also the recommendations in Ref.~\cite{Abdallah:2020pec}. In passing, we note that we have also investigated here the impact of jet-matching / merging together with the different options for the dynamical scale choices offered by {\tt MadGraph5\_aMC@NLO}. The effect is at the percent level, and as such can be ignored if we take into account that ATLAS reports 10\% of uncertainty in the signal modeling, and 10 \% of uncertainty stemming from the parton distribution functions (PDF).

\begin{figure}[h!]
 \centering
 \subfloat[]{\includegraphics[width=0.43\textwidth]{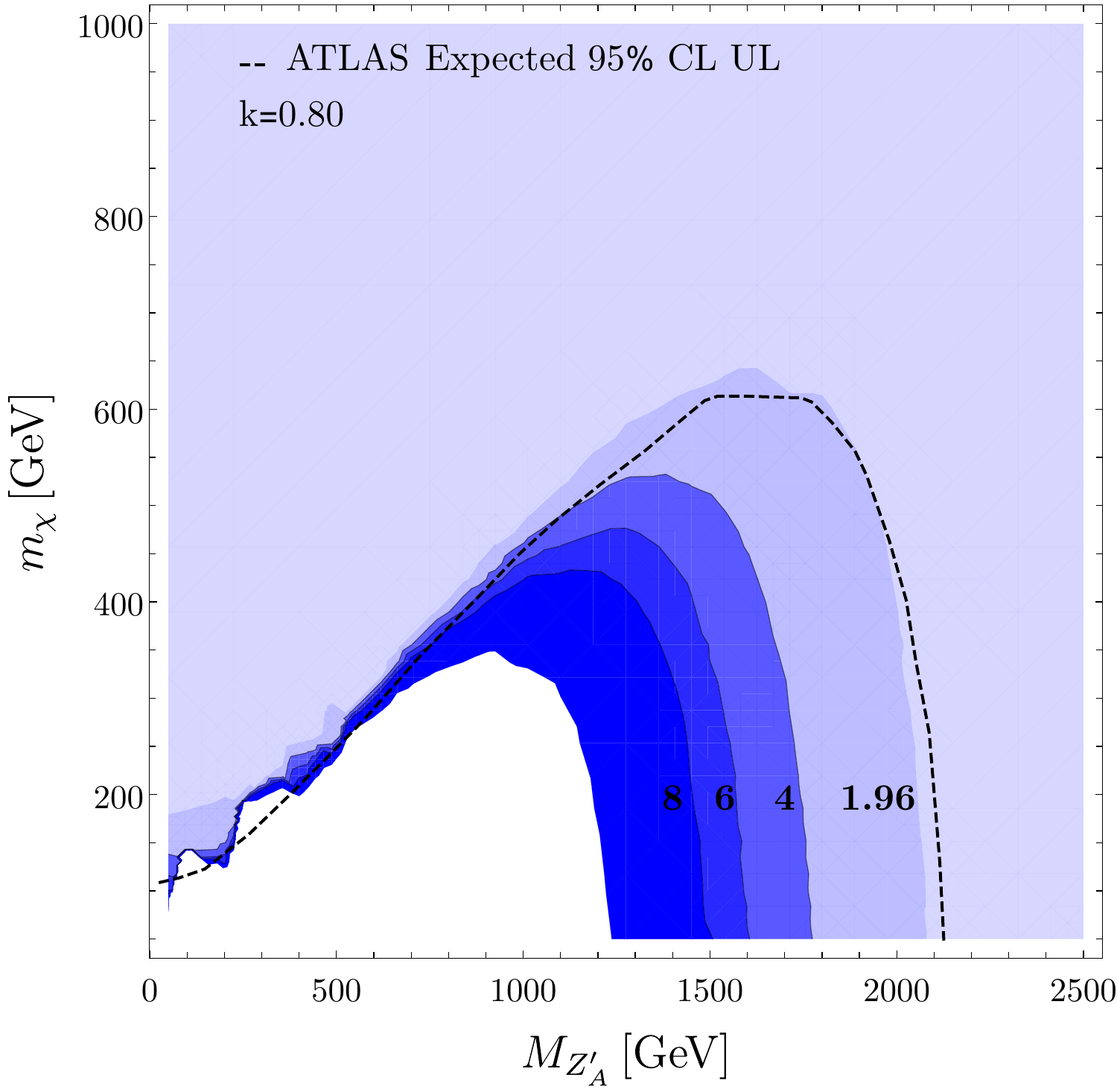}} \hspace{3mm}
 \subfloat[]{\includegraphics[width=0.48\textwidth]{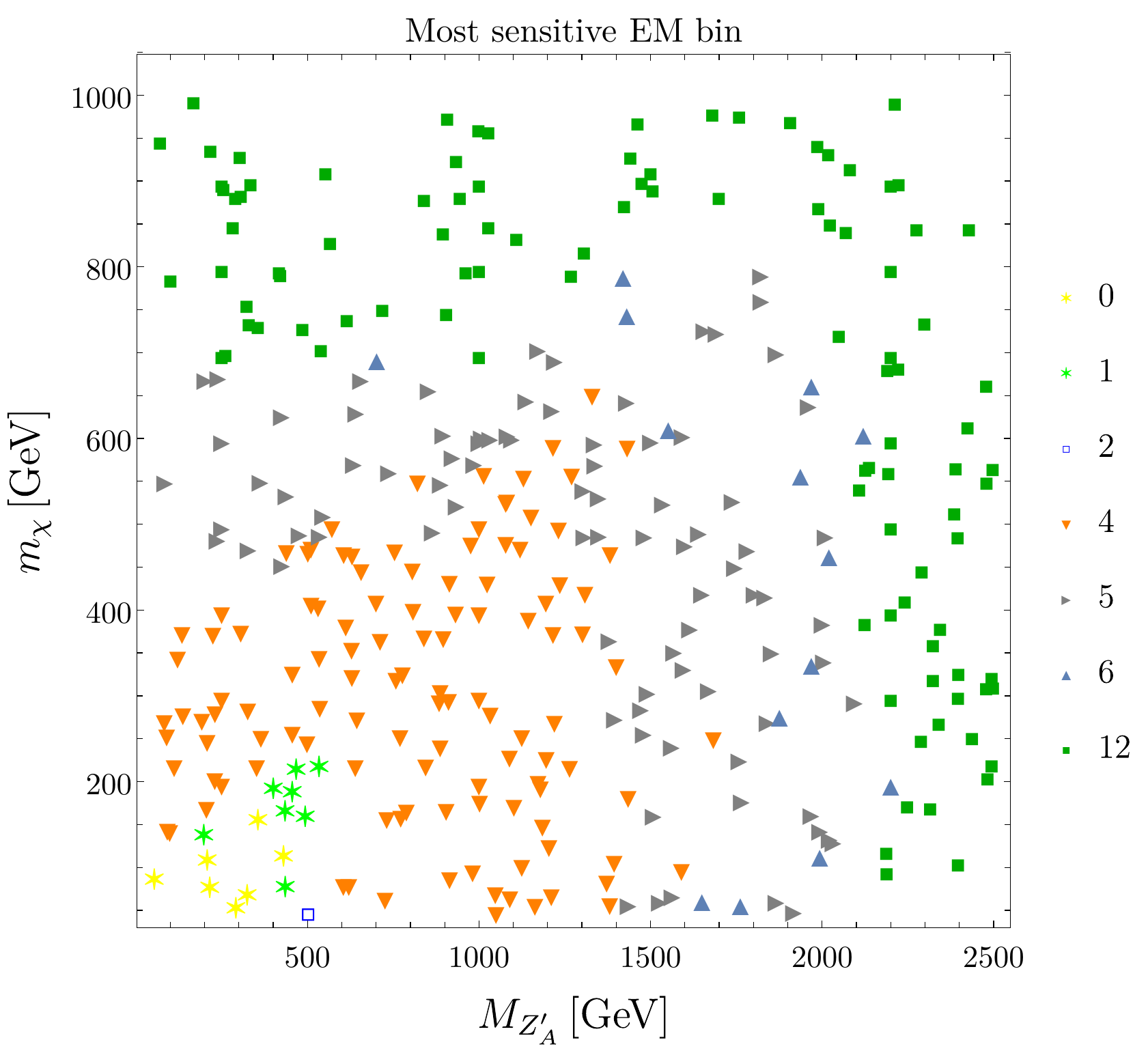}}
 \caption{\small (left) 95 \% C.L exclusion contours for the significance $z={2,4,6,8}$ and the exclusive selection in Table~\ref{tab:EM}, for an axial-vector mediator, with $g_{\chi}=1$ and $g_q=0.25$.  (right) Most sensitive exclusive bin for all the reference points considered in the $(M_{Z'},m_{\chi})$ plane, for an axial-vector ($b$) mediator with $g_{\chi}=1$ and $g_q=0.25$, where the sensitivity is estimated in the Gaussian limit.}
 \label{fig:AV_exclusions}
\end{figure}

\begin{figure}[h!]
 \centering
 \subfloat[]{\includegraphics[width=0.43\textwidth]{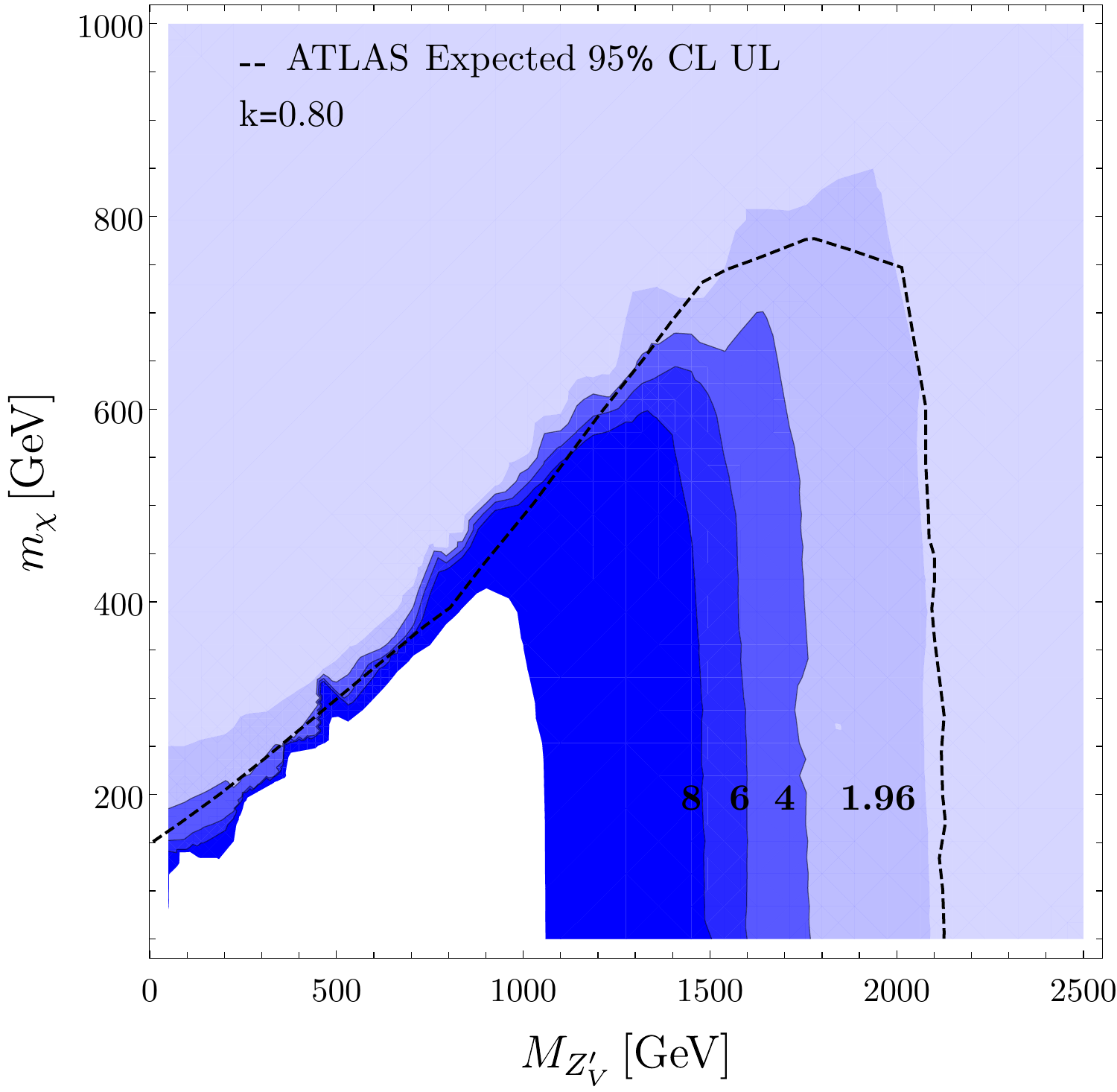}} \hspace{3mm}
 \subfloat[]{\includegraphics[width=0.48\textwidth]{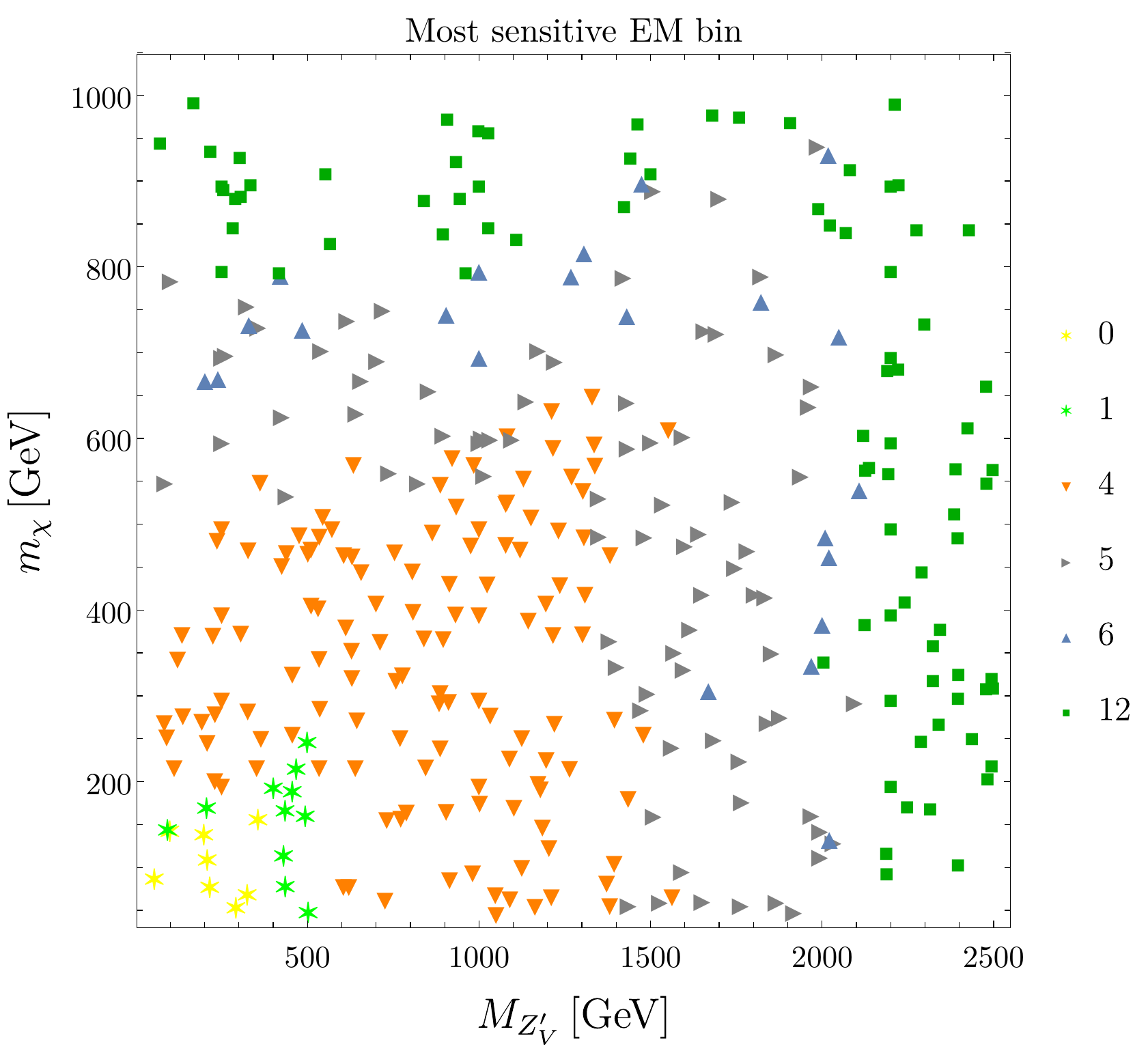}}
 \caption{\small Same as figure~\ref{fig:AV_exclusions}, for the vector mediator case.}
 \label{fig:VEC_exclusions}
\end{figure}

We then proceed to derive the 95 \% C.L exclusion contours for the axial mediator model, which we present in the left panel of figure~\ref{fig:AV_exclusions}. We have considered here 366 points in the 2D mass plane. We display the contours where the significance $z=s_i / \sqrt{b_i}$  is {2,4,6} and 8, together with the ATLAS 95 \% C.L expected exclusion (dashed black line). 
Considering that the latter is derived from a different event selection than the former, the agreement is striking. We also provide, in the right panel of figure~\ref{fig:AV_exclusions} the most sensitive EM signal region for all of our scanned points. This is an interesting plot for two reasons. First and foremost, it provides important insight on the impact of each signal region, and can help to better understand the limits and suggest improvements to the search. Second, this is a relatively simple piece of information to provide, and that would be helpful as now the published table with the 95 \% UL on the signal rates could be directly employed. For completeness we present the analogous of figure~\ref{fig:AV_exclusions} for the vector mediator case, in figure~\ref{fig:VEC_exclusions}. We see that the situation is completely analogous, except that the excluded regions do not fully due to the difference in cross sections for each case.

From these two figures we then conclude that the ATLAS mono-jet search, in spite of lacking some detailed validation material, has been reproduced with an exceptionally good agreement, and hence we will consider the implementation of this study in  $Z'$-explorer 2.0 as validated, and we make the code publicly available on GitHub~\cite{github}.


\section{Numerical results}
\label{sec:resu}
In this section we present several plots obtained with $Z'$-explorer 2.0, which serve as an illustration of its capabilities. For each example, we will present the exclusions in the two-dimensional $M_{Z'} - m_{\chi}$ plane, and we will adopt the following convention to display our results. The most sensitive channel in a given point in the mass plane is indicated by its \emph{shape} (stars, circles, diamonds, etc). If a given point is allowed (i.e. not excluded) then the shape is shown in a color other than black, while excluded points have a shape in black.

We will start first working directly on the parametrization of equation~\ref{eq:Zplag}, where each individual coupling is considered a free parameter. Within this framework, we will restrict ourselves to a few slices of the parameter space, stressing the non-trivial interplay among the different search channels. In second place, we will consider a complete model, a string inspired Stückelberg portal, where there are non trivial correlations among the different couplings.

\subsection{Simplified Model}

\begin{figure}[h!]
 \centering
 \subfloat[]{\includegraphics[width=0.48\textwidth]{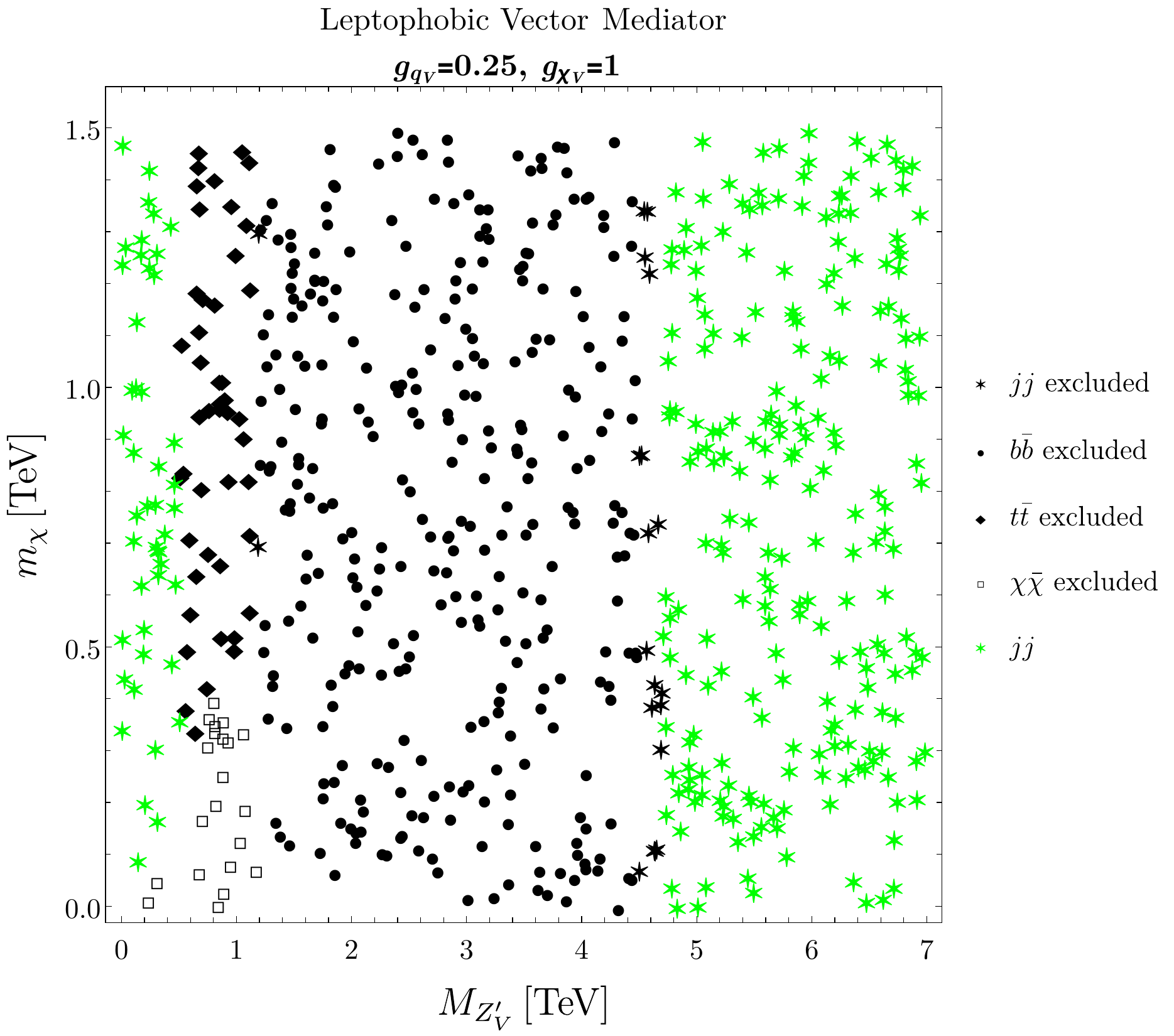}} \hspace{3mm}
 \subfloat[]{\includegraphics[width=0.48\textwidth]{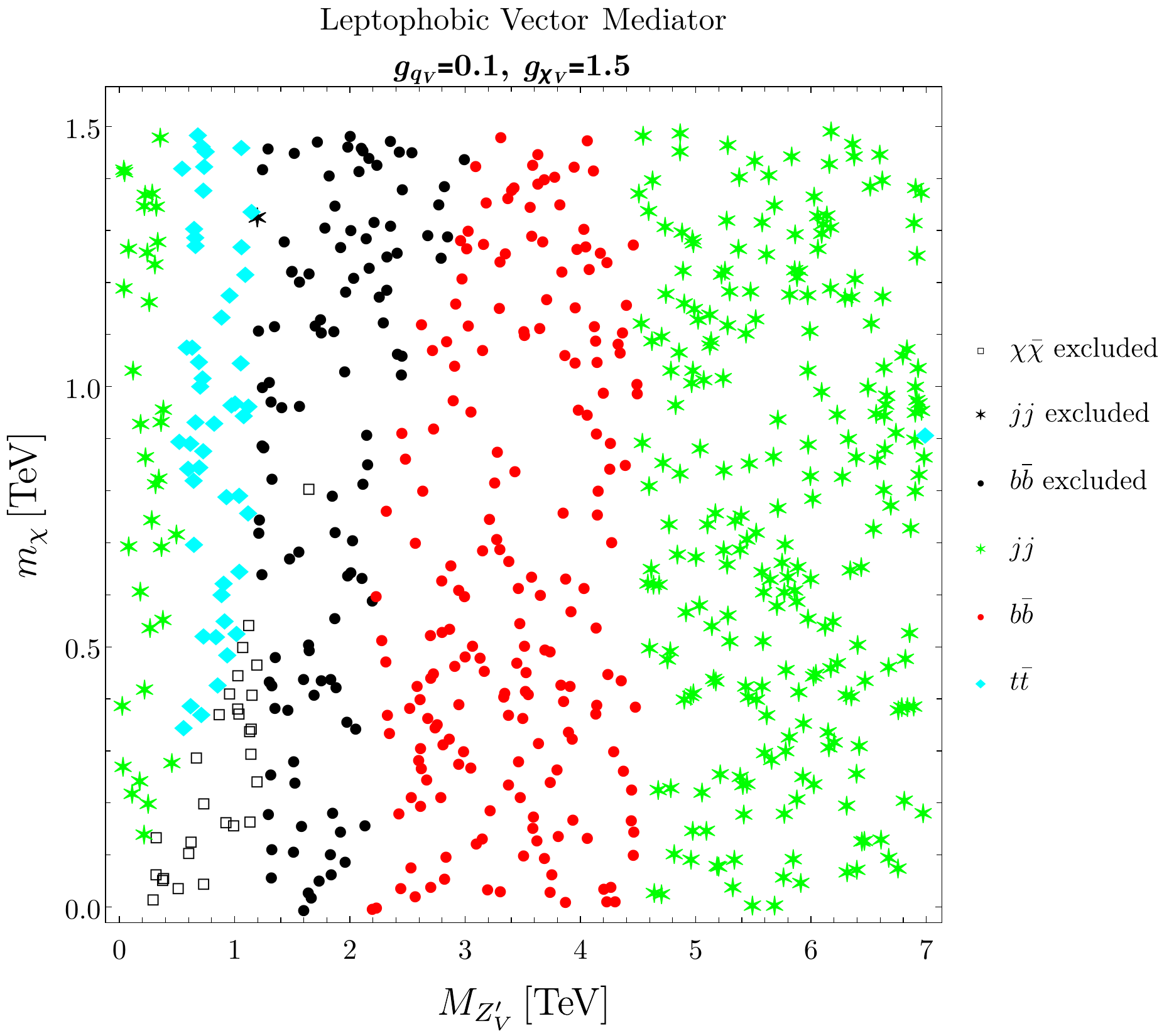}}
 \caption{\small Excluded parameter space and most sensitive channel for the case where the dark matter couples with vector couplings, for two leptophobic $g_l=0$ scenarios, fixing $g_q=0.25$ and $g_{\chi}=1$ (left) and $g_{\chi}=1.5$ (right). The shape of each point shows the most sensitive channel, while a black (colored) point indicates if it is excluded (allowed).}
\label{fig:excl1}
\end{figure}

We start first by examining the benchmark couplings of $g_q=0.25, g_{\chi}=1, g_{l}=0$ used by the ATLAS monojet-study, and for comparison we also include a slightly altered version with $g_{\chi} = 1.5$. These two \emph{leptophobic} cases are shown in the left and right panel of figure~\ref{fig:excl1}.

From the figure one clearly sees the impact of the mono-jet search. Clearly when the dark matter channel is kinematically closed, the situation is relatively simple: depending on the specific mass, a hadronic channel (either $jj$, $b\bar{b}$ or $t \bar{t}$) provide the strongest constraints. The excluded masses range then from 0.5 to 4.5 TeV approximately, with the upper value being due to a reduction of the production cross section, while the lower value is due to the fact that $Z'$-explorer only included information on the visible channels from 400 GeV. This limitation is due to the fact that for lower masses, the inclusive $H_T$ trigger is less efficient due to the overwhelming QCD multi-jet background~\footnote{Nonetheless, there are efforts to reduce this threshold and extend the analysis into lower masses, see e.g.~\cite{Shimmin:2016vlc,CMS:2017dcz,ATLAS:2018hbc}.}. The mono-jet channel is only relevant for the on-shell region, but as soon as it is open it dominates the exclusion, up to a value of $M_{Z'}$ of about 1 (1.5) TeV in the left (right) panel. Comparing the shape of the mono-jet exclusion among both panels, we see that it is relevant for the case where $g_{\chi}$ dominates over $g_q$.

\begin{figure}[h!]
 \centering
 \subfloat[]{\includegraphics[width=0.48\textwidth]{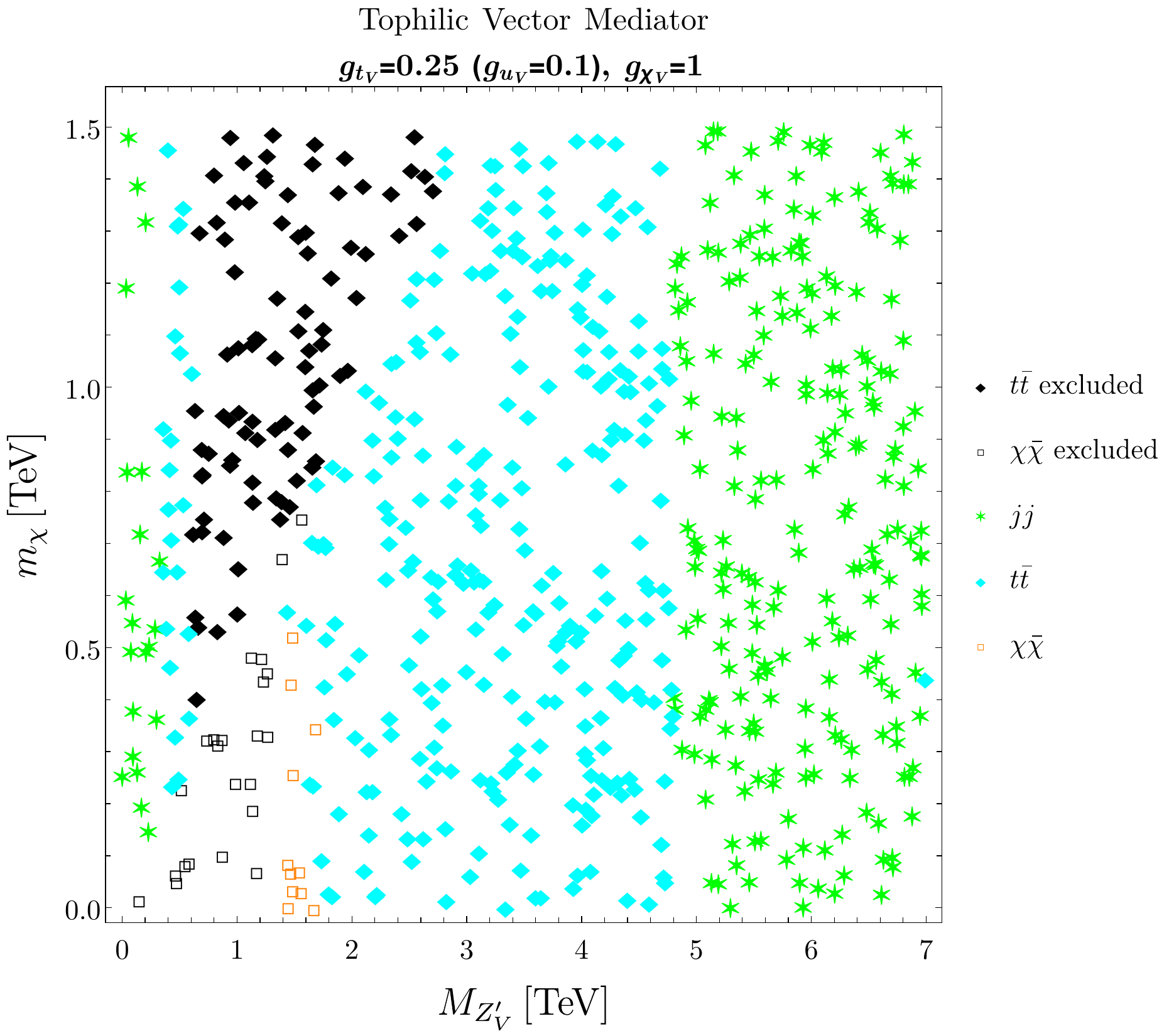}} \hspace{3mm}
 \subfloat[]{\includegraphics[width=0.48\textwidth]{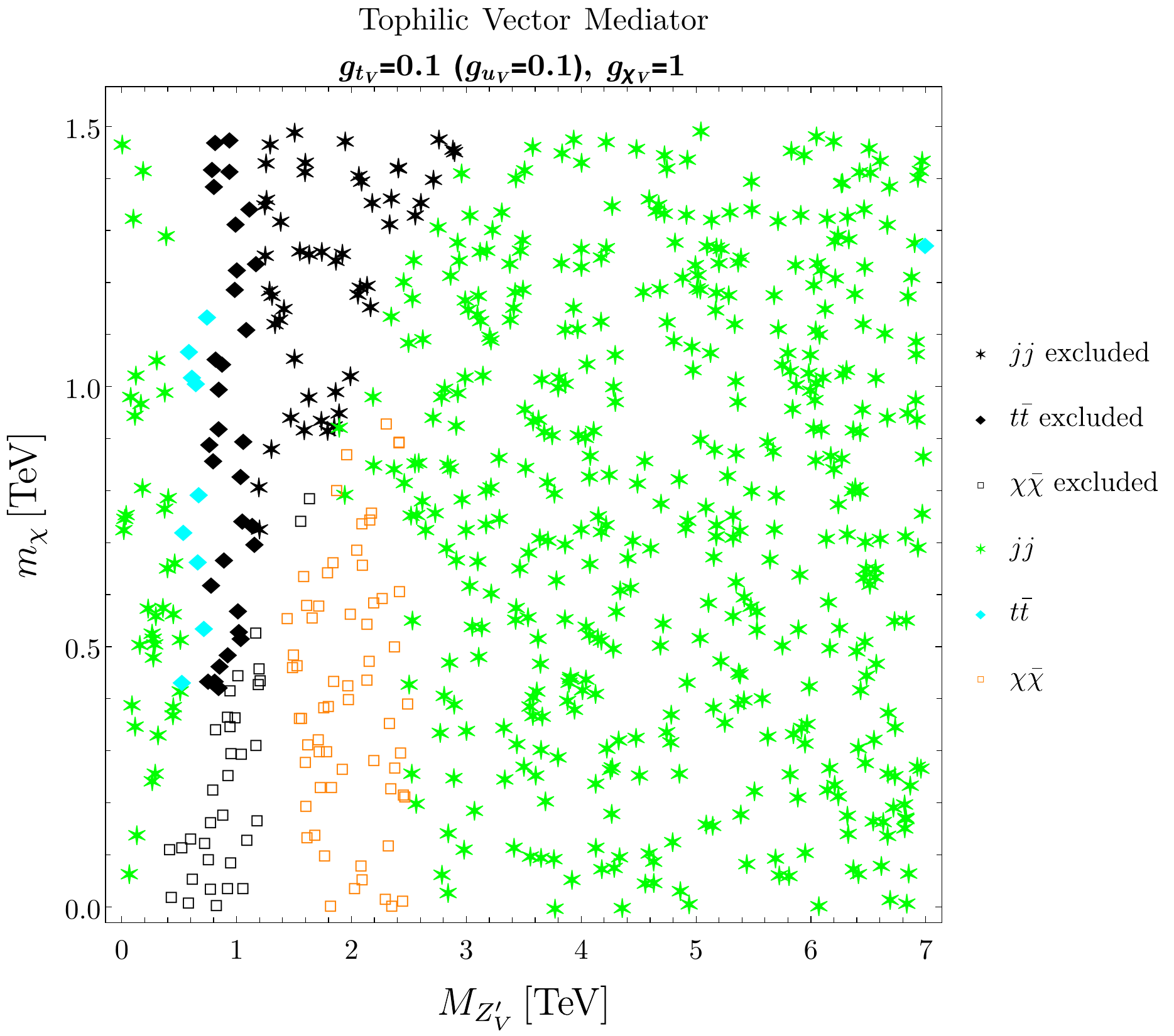}}
 \caption{\small Excluded parameter space and most sensitive channel for the case where the dark matter couples with vector couplings, for two top-philic scenarios (and also leptophobic), fixing $g_u=0.25$ and $g_{\chi}=1$, with $g_t = 0.25 (0.1)$ in the  left (right) panel and $g_{\chi}=1.5$ (right). }
\label{fig:excl2}
\end{figure}

In second place we consider the case where the $Z'$ decays almost exclusively to the top quark (top-philic $Z'$) while still being leptophobic $g_l = 0$. Note that a non zero $g_{q}$ for $u,c,d,s$ needs to be added in order to produce the $Z'$ at the LHC. For simplicity and to avoid having to consider FCNC bounds, we set consider all fermions coupling vectorially, and set $g_u=0.1$, and set $g_{\chi}=1$. We present in the left (right) panel of figure~\ref{fig:excl2} the different exclusions for a medium (small) $g_t$ couplings: 0.25 (0.1).

As expected, this figure illustrates that when the $Z'$ couples more strongly to tops, the sensitive channels in light di-jet resonance become ineffective, and the mono-jet channel can be the most sensitive one in a large fraction of parameter space. We note that in particular, the mono-jet channel can be the most sensitive up to masses of about 2.5 TeV (obviously having the $Z' \to \chi \chi$ channel open).

\begin{figure}[h!]
 \centering
 \subfloat[]{\includegraphics[width=0.48\textwidth]{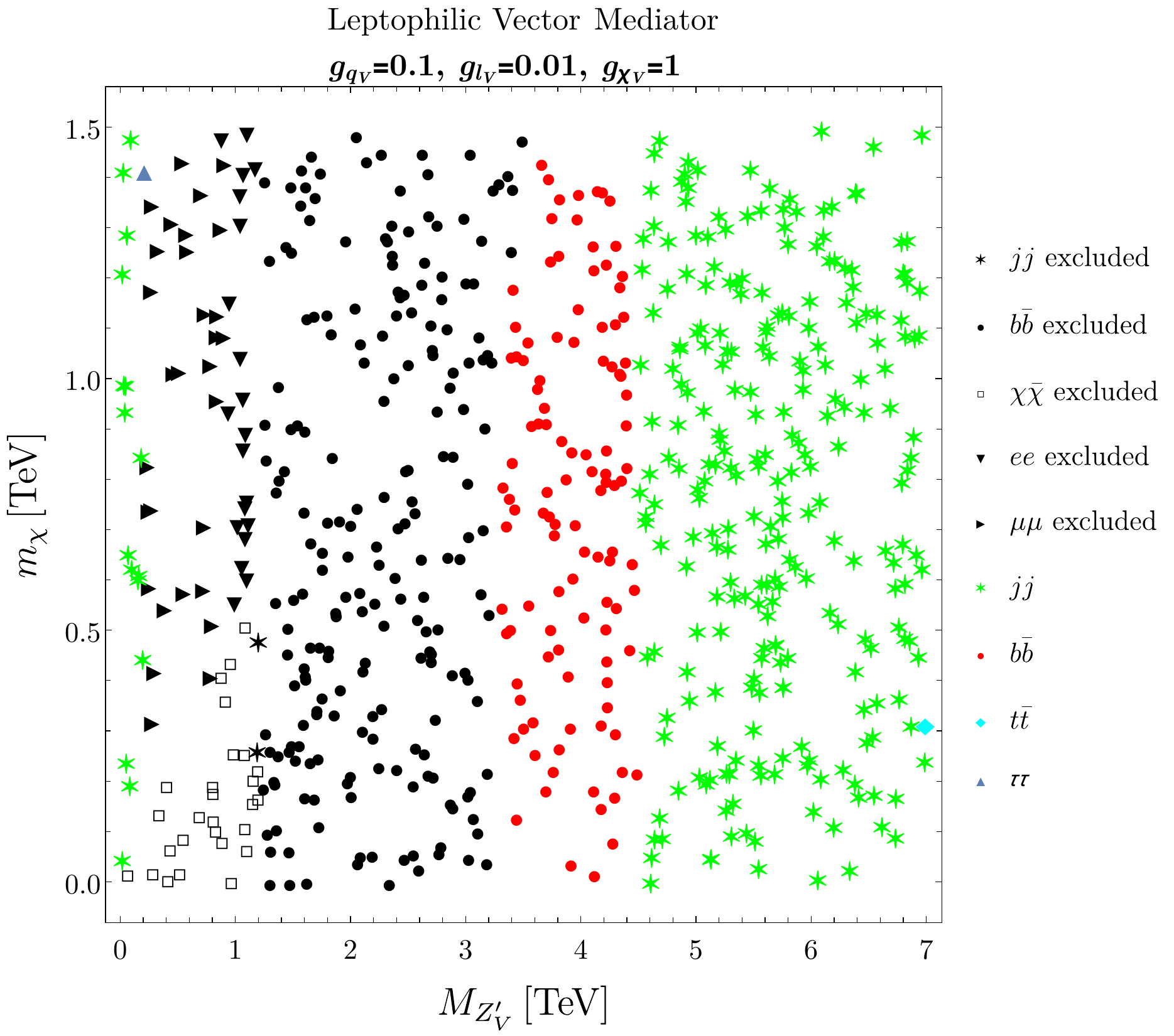}} \hspace{3mm}
 \subfloat[]{\includegraphics[width=0.48\textwidth]{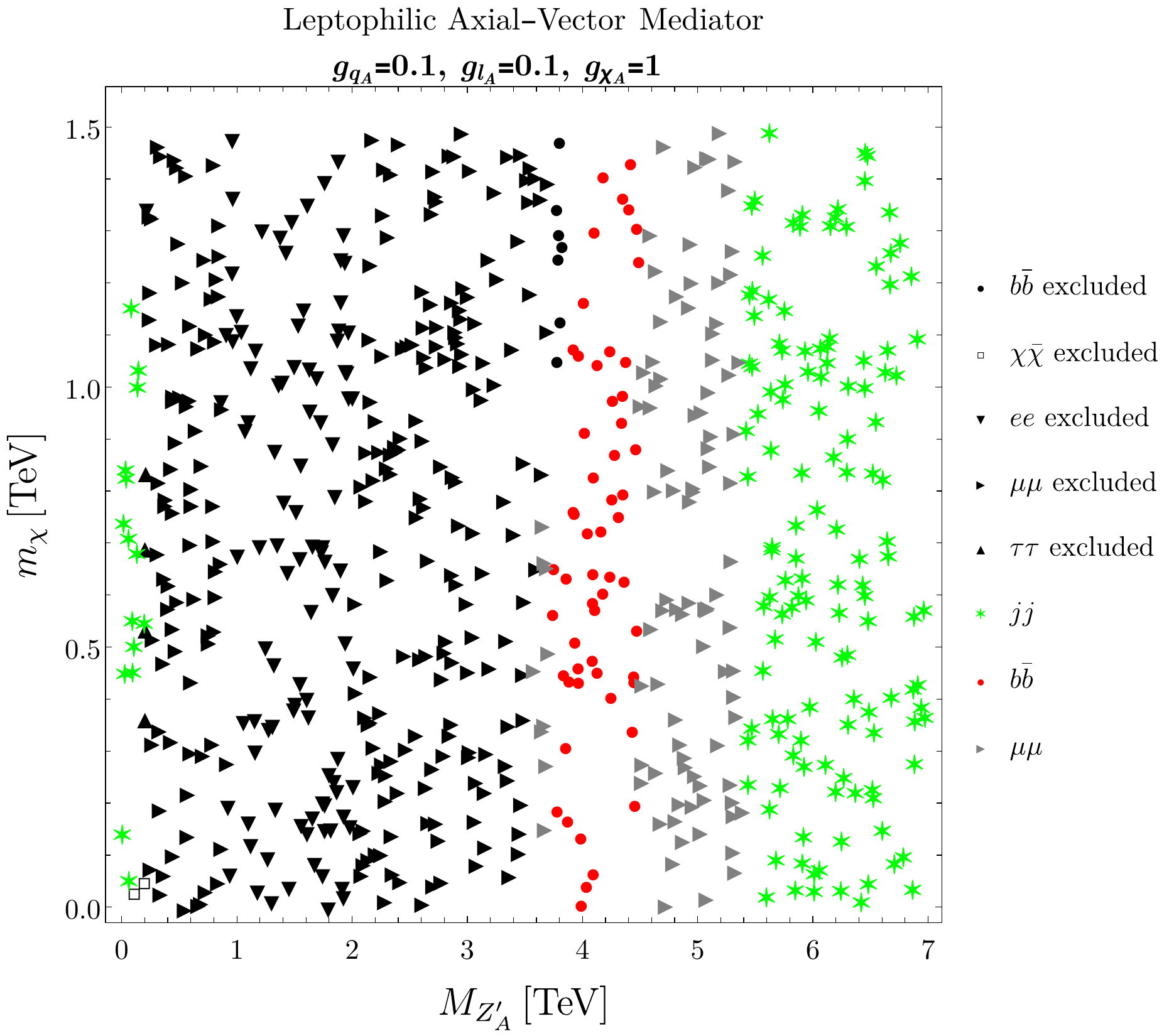}}
 \caption{\small Same as figure~\ref{fig:excl1} and ~\ref{fig:excl2}, yet allowing for a non-zero coupling to leptons. We set $g_q=0.1$ and $g_{\chi}=1$, and show the results for $g_l = 0.01 (0.1)$ in the left (right) panel, for the vector (axial-vector) case.}
\label{fig:excl3}
\end{figure}

For the next example, we drop the assumption of the $Z'$ being leptophobic, and hence consider both vector and axial vector scenarios, with $g_q = 0.1$, $g_{\chi}=1$ and $g_l = 0.01 (0.1)$ for the vector (axial-vector) case. The results plots are shown in the left and right panel of figure~\ref{fig:excl3}

From the figure we clearly see that the lepton channels are very sensitivie, and if they are open they can dominate the exclusions. For large enough couplings the mono-jet search is much less sensitive than the di-lepton channels, except for the case where $g_l =0.01$, where more points have mono-jet as they most sensitive probe.

\begin{figure}[h!]
 \centering
 \subfloat[]{\includegraphics[width=0.45\textwidth]{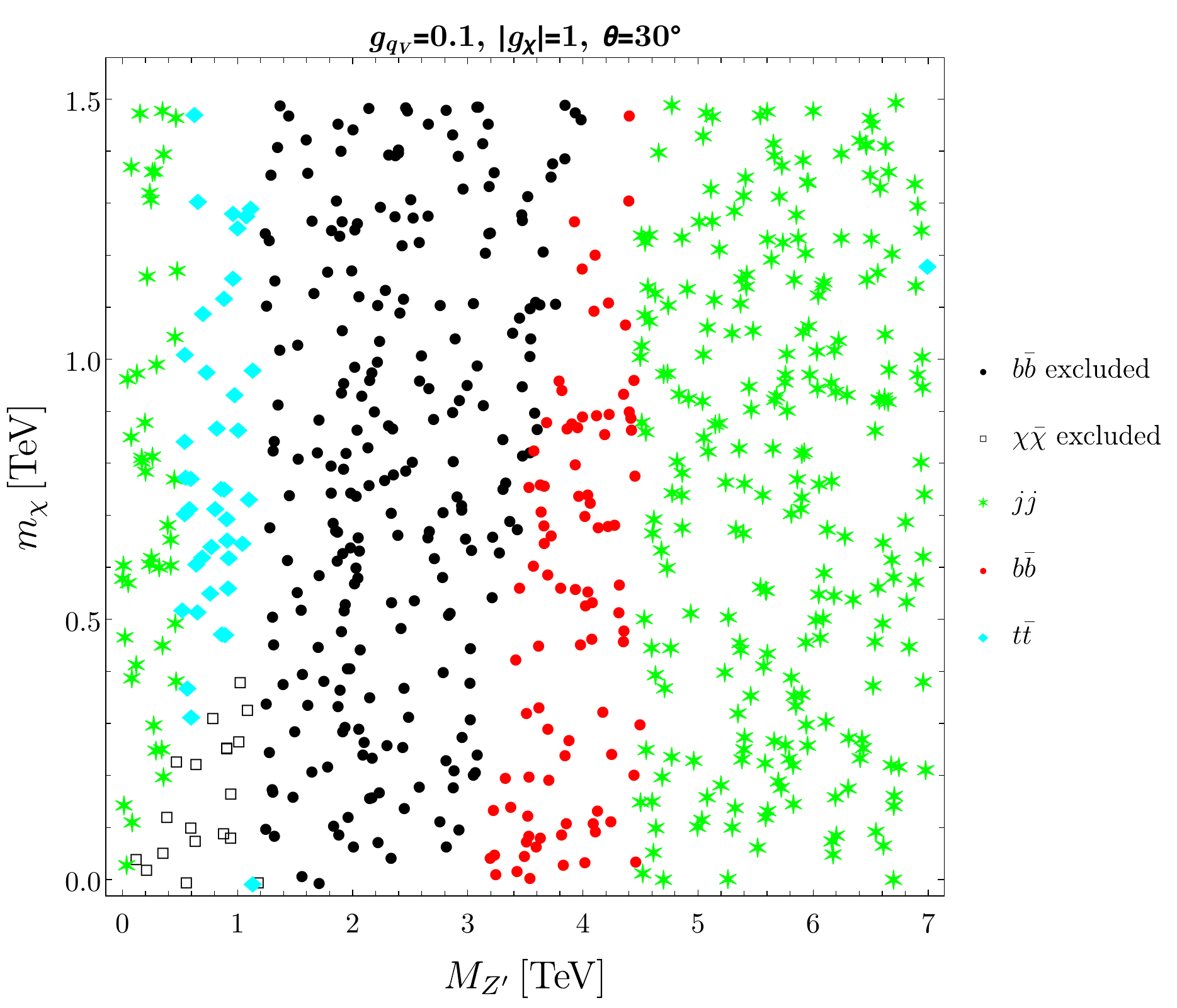}} \hspace{3mm}
 \subfloat[]{\includegraphics[width=0.45\textwidth]{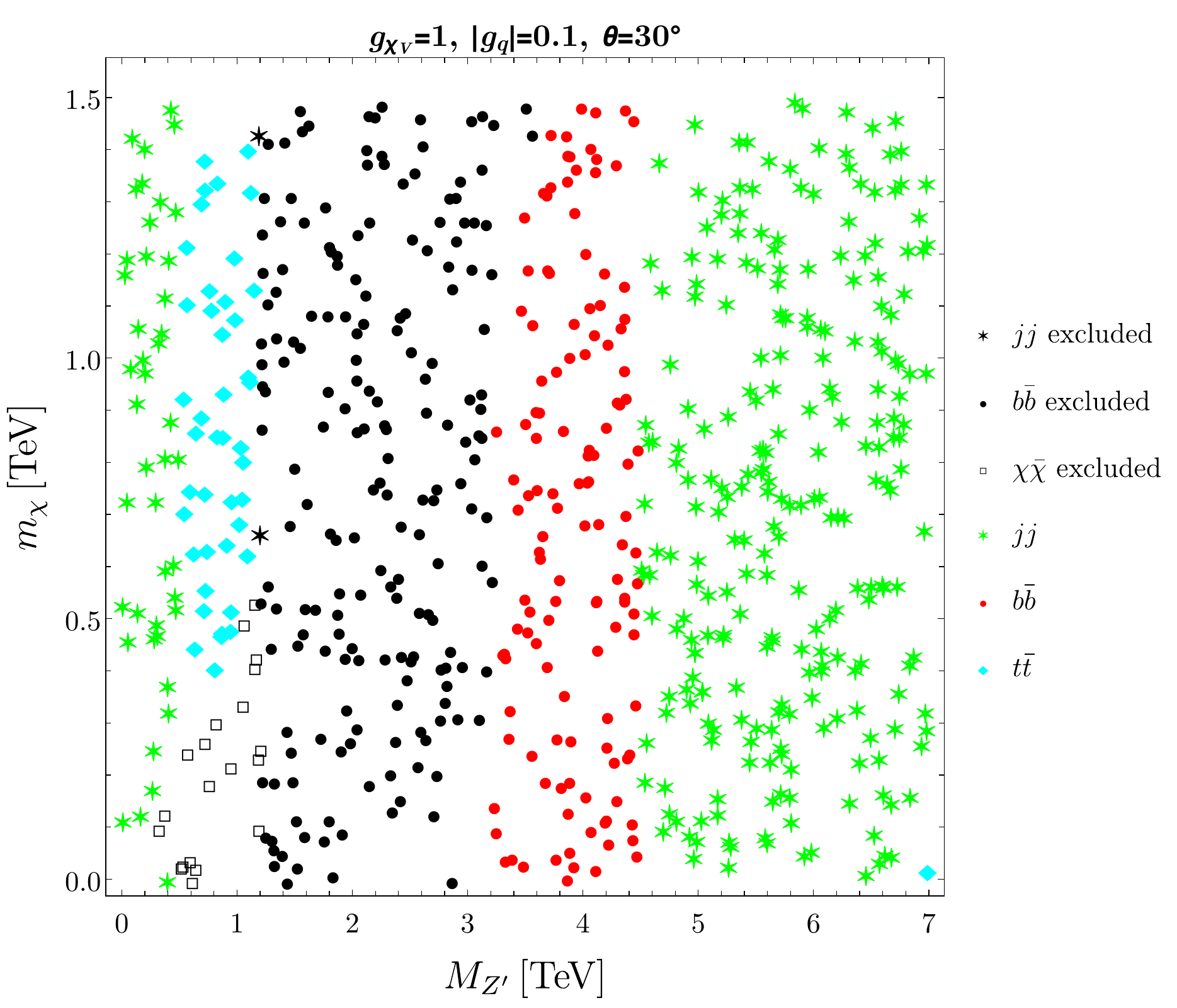}}  
 \vspace{1mm}
 \subfloat[]{\includegraphics[width=0.45\textwidth]{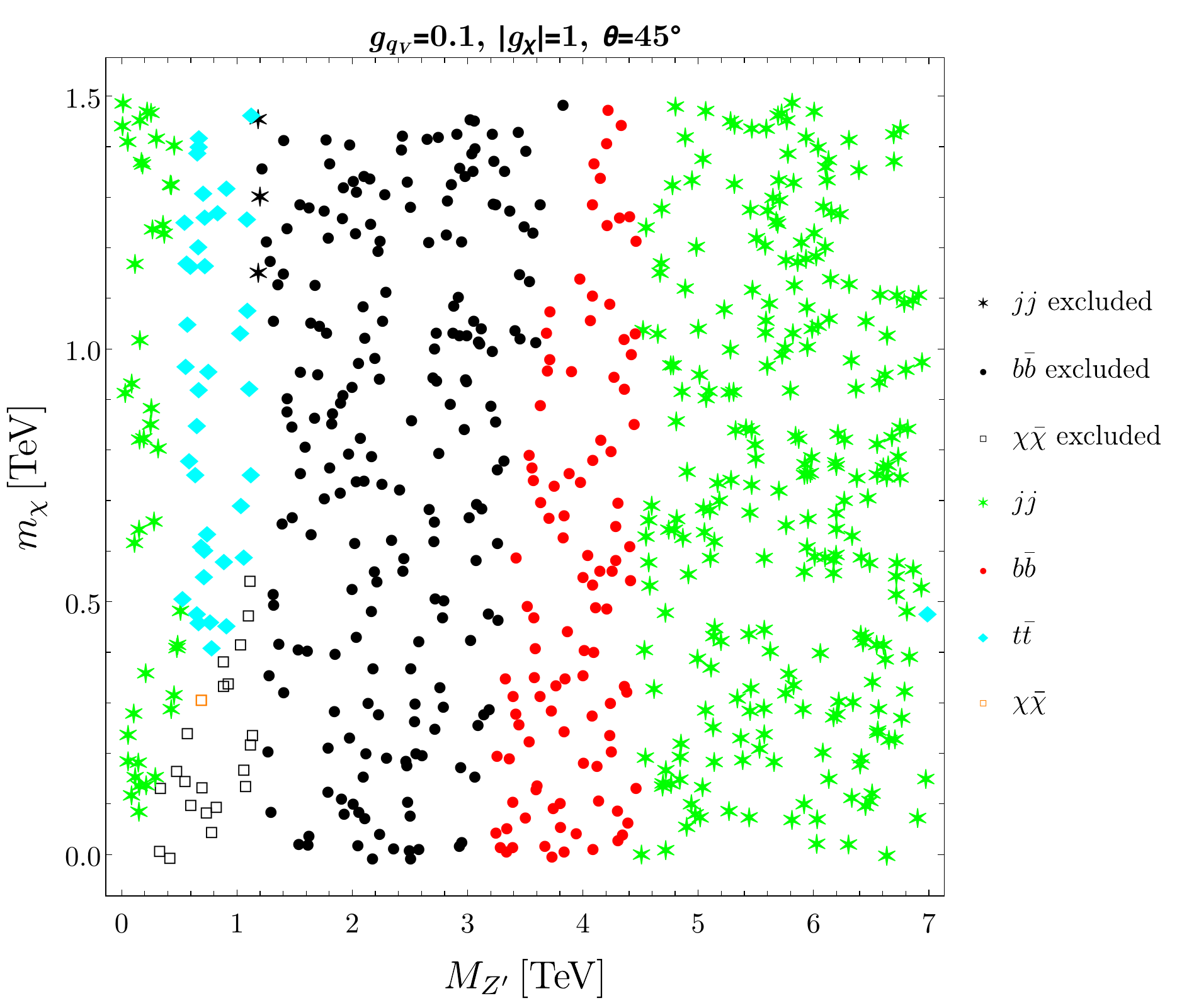}} \hspace{3mm}
 \subfloat[]{\includegraphics[width=0.45\textwidth]{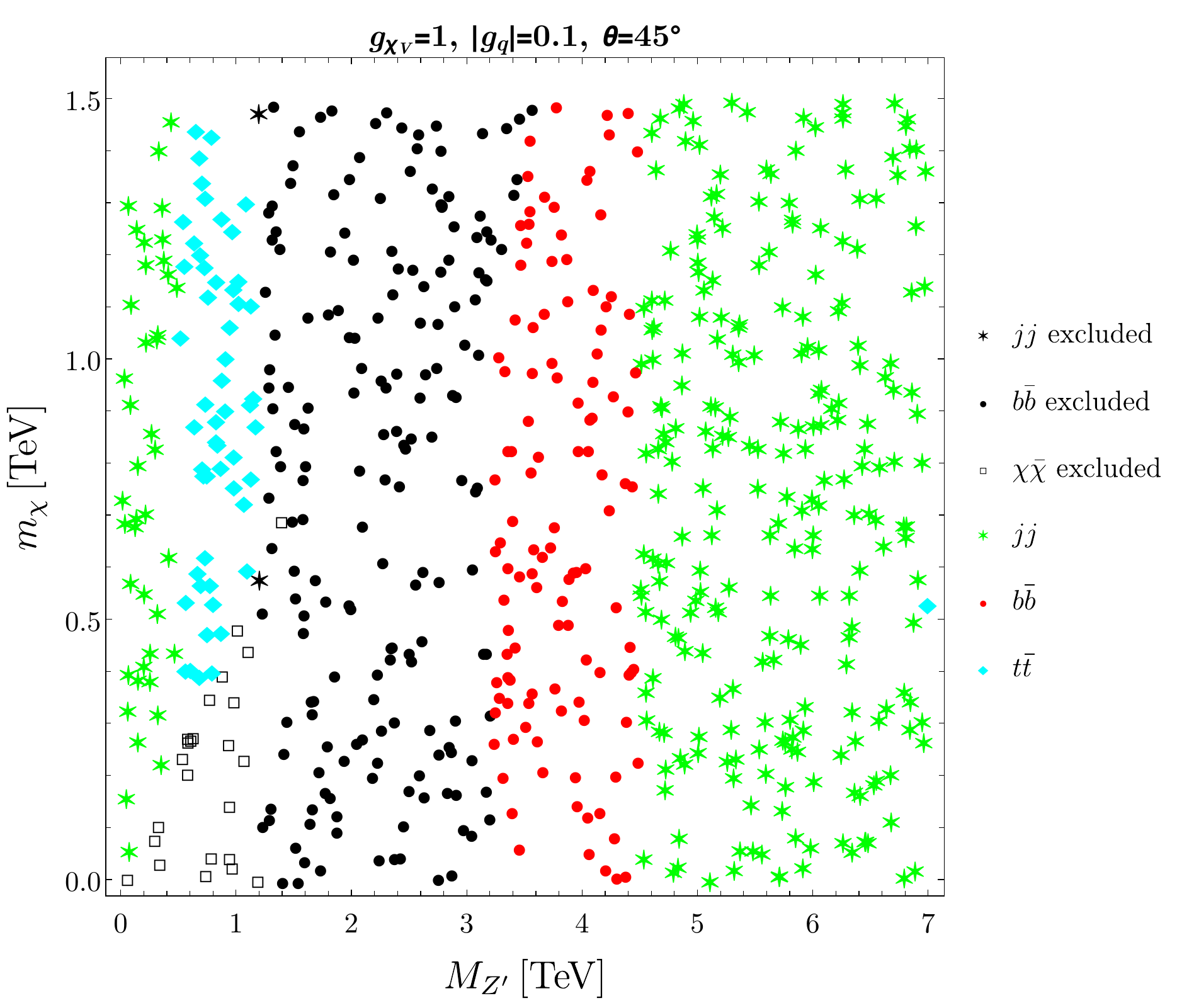}} 
   \vspace{1mm}
 \subfloat[]{\includegraphics[width=0.45\textwidth]{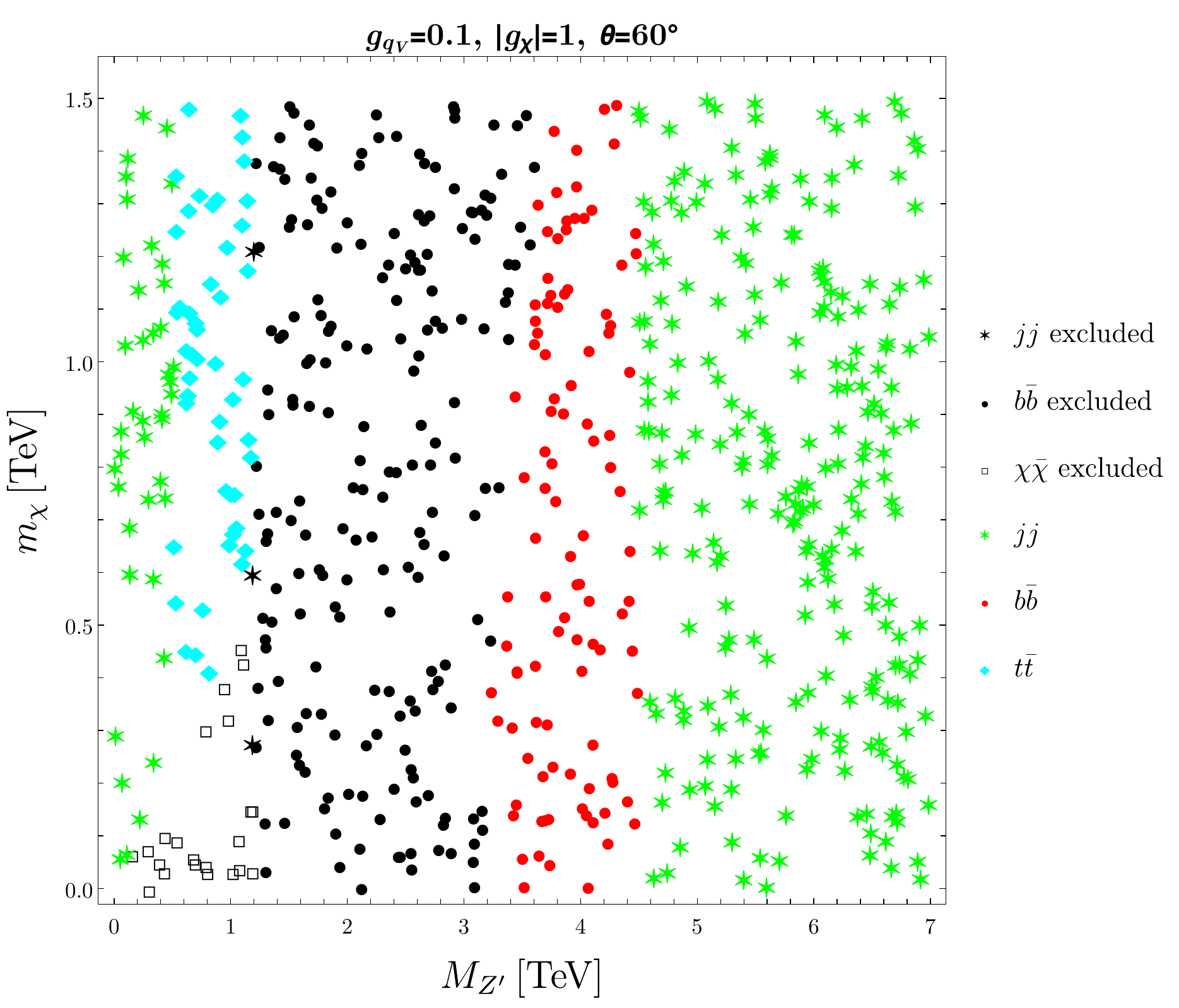}} \hspace{3mm}
 \subfloat[]{\includegraphics[width=0.45\textwidth]{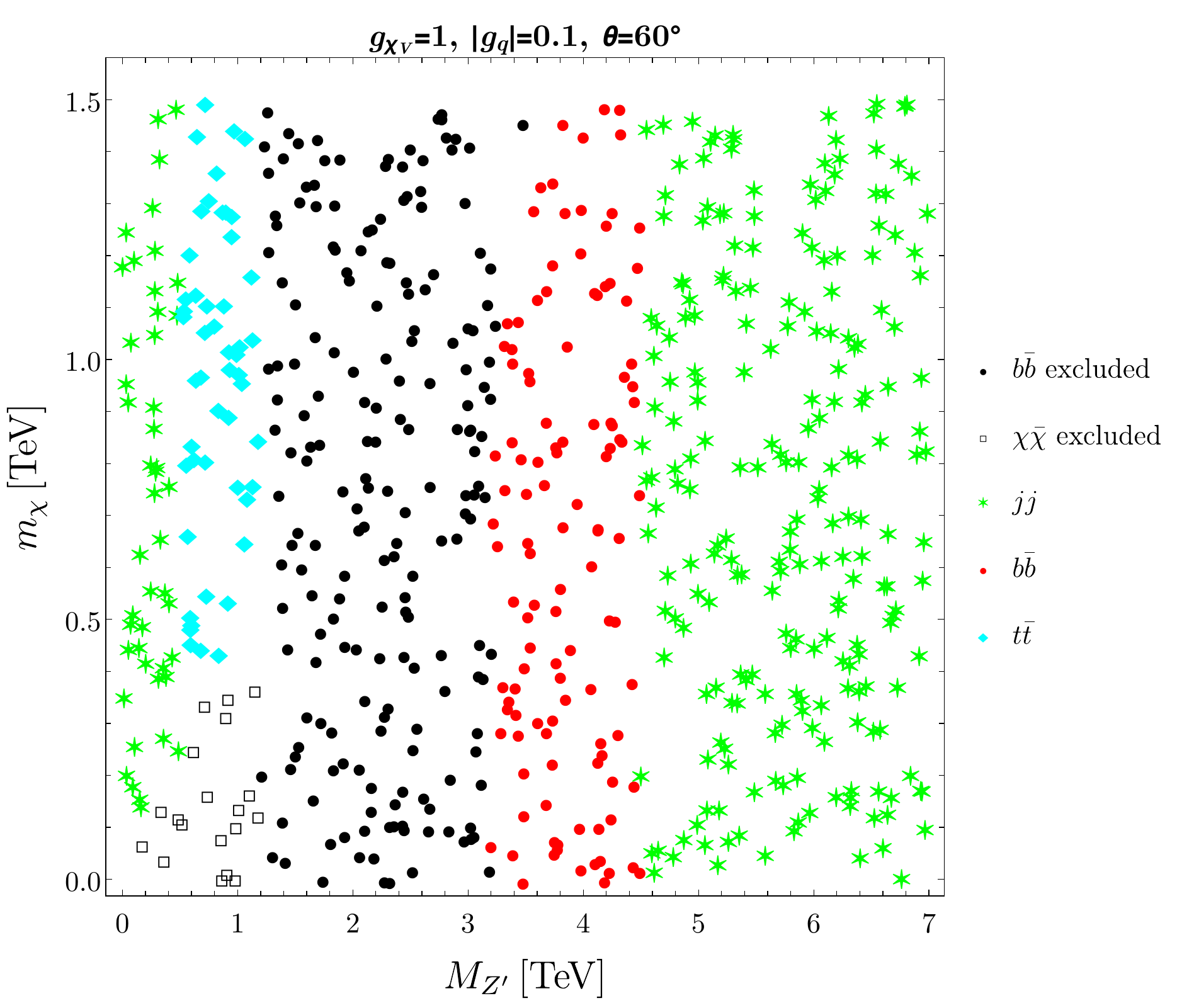}}  
 \vspace{-1mm}
 \caption{\small In the left (right), excluded parameter space and most sensitive channel for the case where the SM quarks (dark matter) couple (couples) with both vector and axial-vector couplings, fixing $g_{q_V} = 0.1$ and $|g_{\chi}| = 1$. The upper, middle and lower panels show the case where the phase $\theta$ between the vector and axial-vector case are 30$^{\circ}$, 45$^{\circ}$ and 60$^{\circ}$. }
\label{fig:excl4}
\end{figure}

Next, we examine dropping the assumption that the $Z'$ must be either a vector or an axial-vector. We then consider a generic mixing angle among both possibilities, first for the quarks, and then for the dark matter $\chi$, both in figure~\ref{fig:excl4}. In each figure we fix the absolute value of $g_\chi =1$ and $g_{q} = 0.1$. We fix also one of the couplings in vector type, and for the other coupling, we consider three phases: 30$^{\circ}$, 45$^{\circ}$ and 60$^{\circ}$ ($\theta=0^{\circ}$ corresponds to a pure axial-vector case, and $\theta=90^{\circ}$ to a pure vector case).  What see in the figures is that the exclusions are rather similar, only presenting small variations. We thus conclude that while one can in principle be sensitive to the phase, in practical terms the variation with the angle is fairly small and can be hard to extract, from a potential signal, the $\theta$ phase.

\subsection{St\"uckelberg portal from intersecting D6 branes}

\begin{table}[]
	\begin{center}
		\begin{tabular}{||c||c|c|c|c|c||}
			\hline
			Matter field	& $Q_A$ & $Q_B$  & $Q_C$  & $Q_D$  & $Y$ \\
			\hline
			\hline
			$Q_L$	& 1  & -1 &  0 &  0 &  1/6 \\
			$q_L$	& 1  &  1 &  0 &  0 &  1/6 \\
			$U_R$	& -1 &  0 &  1 &  0 & -2/3 \\
			$D_R$	& -1 &  0 & -1 &  0 &  1/3 \\
			\hline
			$L$	    & 0  & -1 &  0 & -1 & -1/2 \\
			$E_R$	& 0  &  0 & -1 &  1 &  1   \\
			$N_R$	& 0  &  0 &  1 &  1 &  0 \\
			\hline
		\end{tabular}
		\caption{\small Charge assignment of the SM matter particles under the four $U(1)$ factors of Ref.~\cite{Ibanez:2001nd}.}
		\label{tab:charges}
	\end{center}
\end{table}

Next we will test the power of the tool $Z'$-explorer with a complete model that involves more complicated relations among the couplings of the $Z'$. This particular model consists of a Stückelberg portal that could arise from intersecting D6 branes that was studied in Refs.~\cite{Feng:2014eja,Feng:2014cla,Lozano:2015vlv}. We will focus on a particular gauge sector given by 
\begin{eqnarray}
SU(3)_c\times SU(2)_L \times U(1)_V^A\times U(1)_V^B\times U(1)_V^C\times U(1)_V^D\times U(1)_h^m\times G_h ,
\end{eqnarray}
where the subscript $V$ and $h$ corresponds to visible and hidden sector respectively, $U(1)_h^m$ are the $m$ abelian gauge factors that only couple to the hidden sector while $G_h$ is the semi-simple part of the hidden gauge group. Within the visible sector there are four abelian gauge factor that couple to visible matter fields. Each of these factors\footnote{These factors arise from the different overlapping branes that intersect while obtaining the SM from type IIA string theory, this realization is the so-called {\textit{Madrid quivers}} that is one of the simplest realistic models of intersecting D6 branes~\cite{Ibanez:2001nd}. A stack of $N$ branes usually hosts a gauge group $U(N)\cong SU(N)\times U(1)$.} has a charge $Q_\alpha$ under which the SM matter particles are charged. The charges of the SM matter particles in terms of the four visible $U(1)$ are listed in Table~\ref{tab:charges}. In order to have an anomaly-free model the quark sector is split in two assignments, being $Q_L=t_L,b_L$ and $q_L=u_L,c_L,d_L,s_L$. If we have a look at Table~\ref{tab:charges} we can see that the different charges are related with different physical properties. $Q_A$ is related with baryon number, $Q_D$ with lepton number while $Q_B$ and $Q_C$ are related with the left and right nature of the matter particle. 

To obtain the SM a combination of the four charges must give rise to the hypercharge, in this case and according to Table~\ref{tab:charges} we have,

\begin{eqnarray}
Q^Y=\frac{1}{6}(Q_A-3Q_C+3Q_D).
\end{eqnarray}
This combination remains massless before electroweak symmetry breaking. The other three $U(1)$ gauge bosons left acquire masses by the Stückelberg mechanism~\cite{Stueckelberg:1938hvi,Stueckelberg:1938zz}. Here, we will assume that only one of these bosons is light enough to produce interesting phenomenology at collider energies while the other two are heavy enough to be decoupled from any effect. So any matter field $\psi_\alpha$ will couple to the lightest $Z'$ boson as a combination of the different charges to the $U(1)$ factors,
\begin{eqnarray}
g_\alpha^{Z'}=a Q_{\alpha A} + b Q_{\alpha B} + c Q_{\alpha C} + d Q_{\alpha D} + \sum_{i=1}^m h_i Q_{\alpha i}^h,
\end{eqnarray}
where $h_i$ are the couplings to the hidden sector. 
Once we know how the $Z'$ couples to matter and also the information of the charge assignment given in Table~\ref{tab:charges} we can derive the actual couplings of the SM particles to the $Z'$. As we mentioned before the quark sector is divided between the first and second generation and the third one, as they are charged differently they will consequently have different couplings. The couplings for the first and second generation of quarks are,
\begin{eqnarray}
g_L^{u,c,d,s}= (a+b),\quad g_R^{u,c}= (-a+c),\quad  g_R^{d,s}= (-a-c).
\end{eqnarray}
The couplings for the top and bottom quarks read,
\begin{eqnarray}
g_L^{t,b}= (a-b),\quad g_R^{t}= (-a+c),\quad g_R^{b}= (-a-c).
\end{eqnarray}
And finally the couplings to the leptons are,
\begin{eqnarray}
g_L^{\ell,\nu}= (-b-d),\quad g_R^{\ell}= (-c+d).
\end{eqnarray}
As we can see in the lepton sector there is no distinctions among families.
For the sake of simplicity we will consider that the dark matter couples to the $Z'$ boson with two different couplings representing the left and right ones, $g_\chi^L$ and $g_\chi^R$.

One of the aspects that makes this kind of portals attractive to explain the dark matter sector is the fact that the SM couplings to the $Z'$ can be quite general. In that way the dark matter particle will couple differently through the $Z'$ to protons and neutrons, since the up and down couplings are different. Thus, this will be translated to non-zero isospin violation, $f_n/f_p$, as the general case. This affects directly to the phenomenology of dark matter since the dark matter direct detection experiments rely on the particle dark matter scattering off nuclei~\cite{Feng:2011vu}.

Once we have set the model and obtained the couplings we can now test it against $Z'$-explorer. In order to scrutinise the parameter space of the model we have chosen different scenarios that are phenomenologically interesting, they are shown in Table~\ref{tab:bpstu}.

\begin{table}[]
	\begin{center}
	\begin{tabular}{|c||cccccc|}
		\hline
	        Scenario	& $a$ & $b$ & $c$ & $d$ & $g_\chi^L$ & $g_\chi^R$ \\
	        	\hline
	1	& 0.07 & 0.00058 & 0.01 & 0.006 & 1.0 & -1.0  \\
	\hline
	2	& -0.025 & -0.005 & 0.005  & 0.025 & 1.0 & 1.0 \\
	\hline
	3	& 0.1 & -0.01 & 0.01 & 0.01 & 1.0  & 1.0 \\
	\hline
	\end{tabular}
	\caption{\small Different values of the couplings for the three phenomenological scenarios to be tested against $Z'$-explorer 2.0.}
	\label{tab:bpstu}
    \end{center}
\end{table}

\begin{figure}[h!]
	\centering
	\subfloat[]{\includegraphics[width=0.45\textwidth]{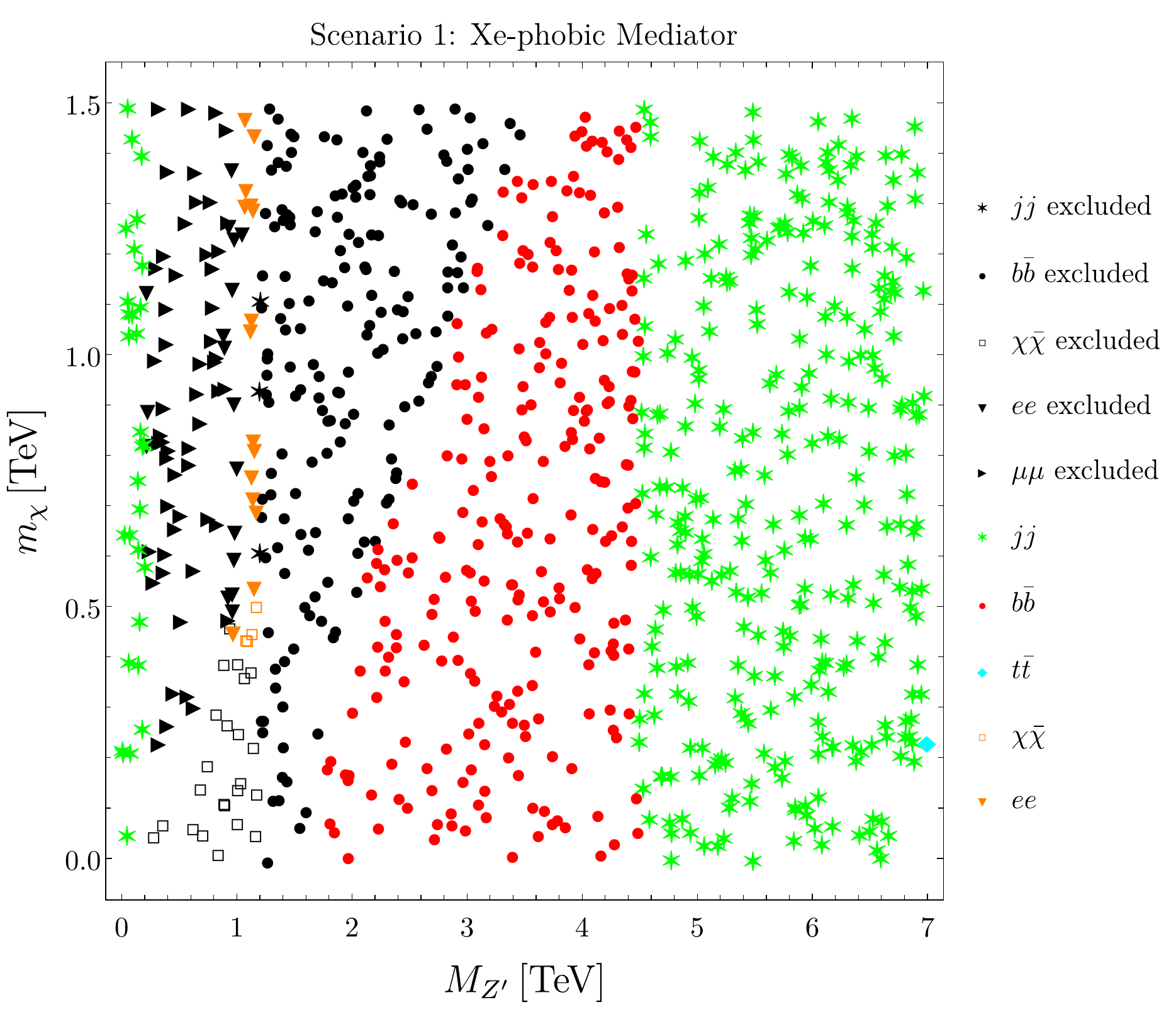}} \hspace{3mm}
	\subfloat[]{\includegraphics[width=0.45\textwidth]{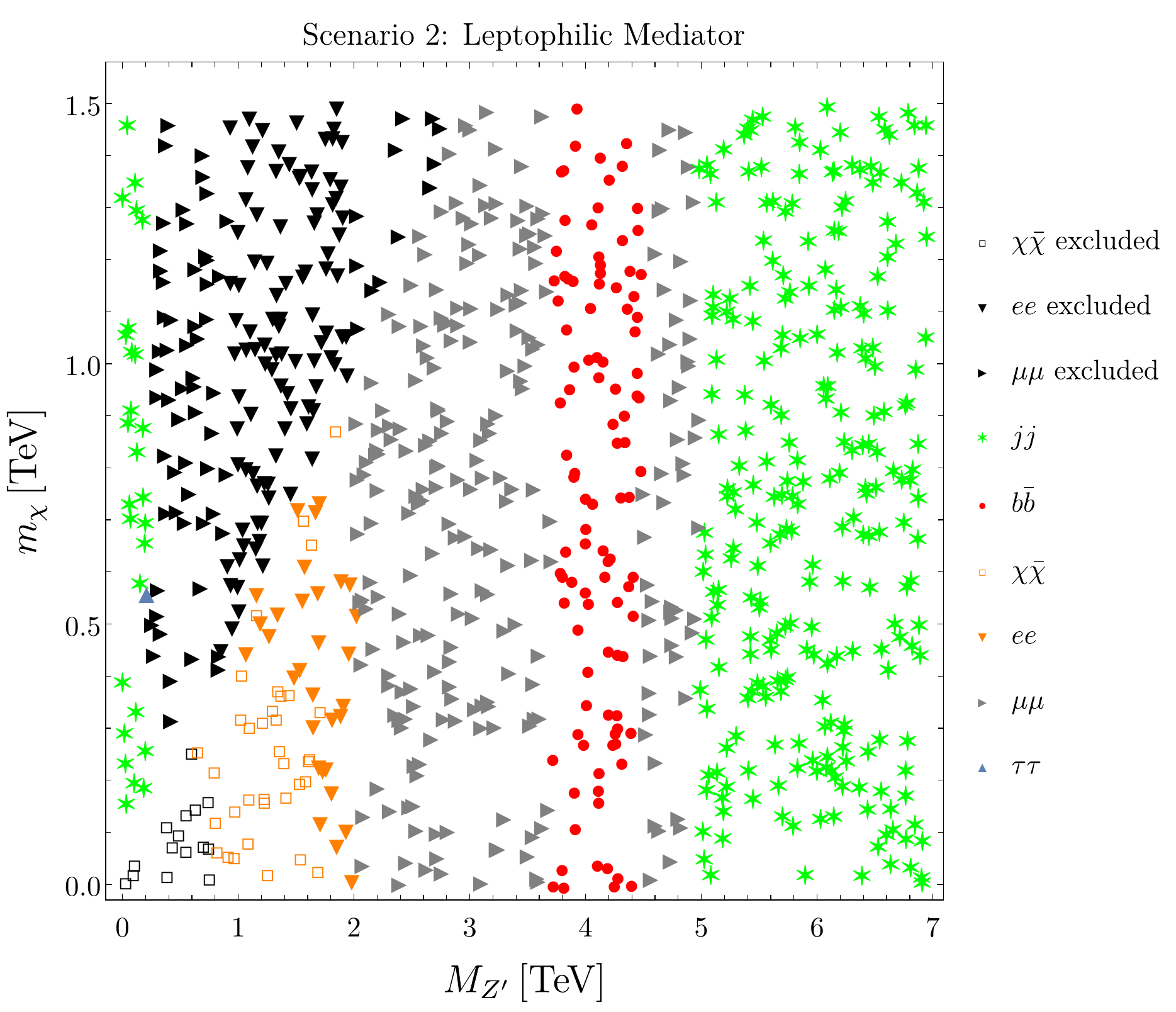}} 
	\subfloat[]{\includegraphics[width=0.45\textwidth]{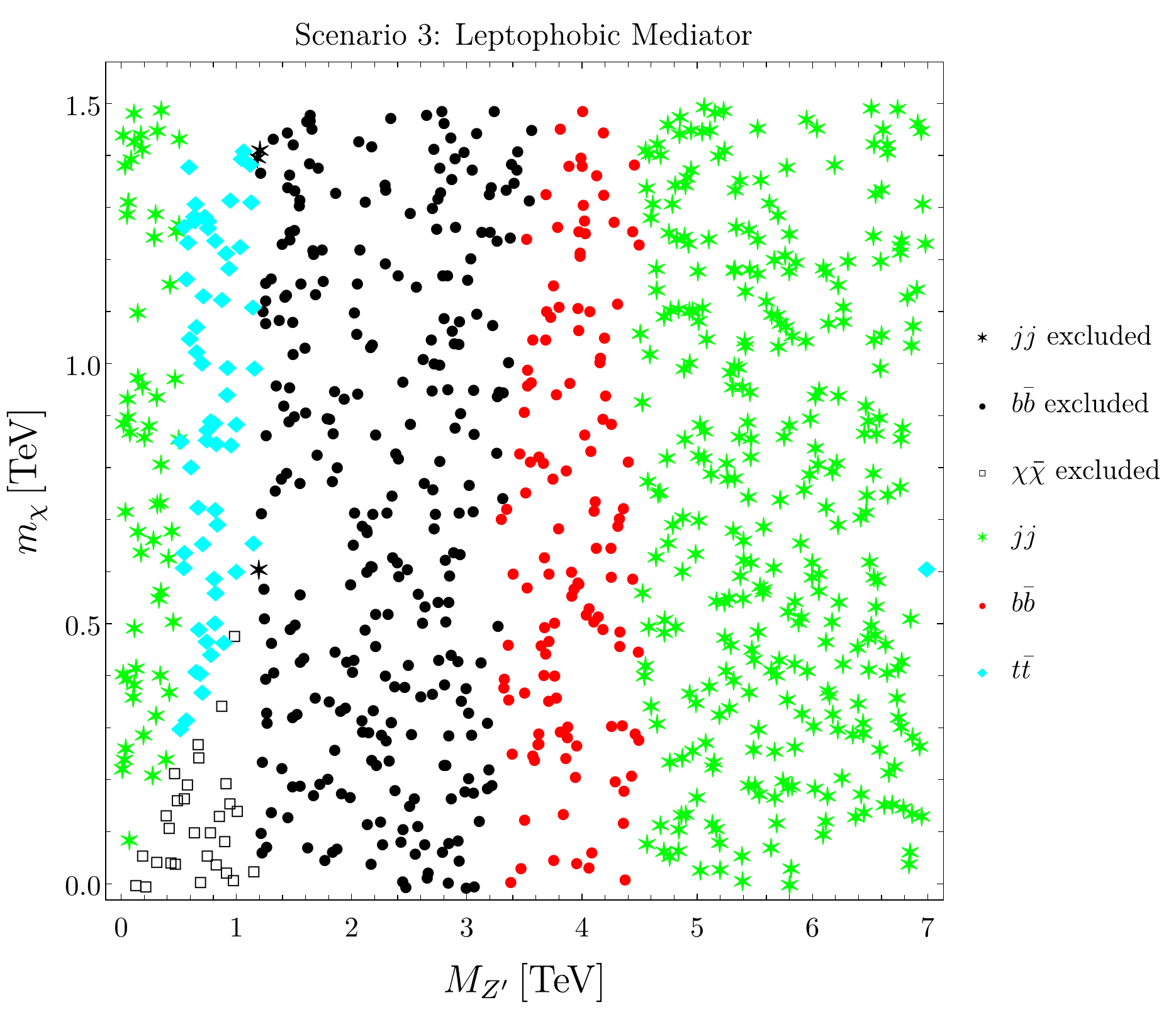}} 
	\caption{\small Excluded parameter space and most sensitive channel for the case of the scenarios listed in Table~\ref{tab:bpstu}. }
	\label{fig:stu_1}
\end{figure}

First of all we examine Scenario 1 where we have an amount of isospin violation that makes Xenon based direct detection of dark matter experiments less sensitive to the $Z'$. This kind of scenarios are usually called Xe-phobic. The amount of isospin violation that one needs to obtain this is $f_n/f_p=-0.7$, so in our construction this is achieved by a combination of the left and right parameters such as $b/c\simeq 0.0588$.\footnote{Details on how the amount of isospin violation is related with the parameters of the model can be found in Ref.~\cite{Lozano:2015vlv}} The parameters that define this scenario are listed in Table~\ref{tab:bpstu} while we scan the mass of the dark matter particle and the $Z'$ in the ranges $m_{\chi}\in (0-1.5)$ TeV and $m_{\chi}\in (0-8)$ TeV. The results are depicted in the left plot of figure.~\ref{fig:stu_1}.

The results for Scenario 1 shows the versatility in searches performed by $Z'$-explorer. First of all we can see that for low dark matter masses and when this decay channel is kinematically allowed the monojet search can exclude up to dark matter masses of $m_\chi\sim 400$ GeV for masses of the $Z'$ up to 1.1 TeV. In the same mass region of the $Z'$ but for larger masses of the dark matter that closes the $Z'\to \chi\chi$ channel, the leptonic searches are the most sensitive ones. The electron and muon searches can exclude all the region between $M_{Z'}=(0.4-1.1)$ TeV. While the exclusion for larger masses of $Z'$ is due to the $b$-quark searches.

In the second scenario we have chosen the parameters in such a way that the b-quark couplings gets shrunk. This makes the coupling to leptons be predominant and that is the reason why we call this scenario leptophilic even if there are some parts of the parameter region where the decay to quarks (including the $b$-quark) can be significant. The monojet channel is the most sensitive one for $Z'$ masses up to 1.6 TeV when the dark matter decay channel is open. However, the power of exclusion only reaches until $M_{Z'}$=1 TeV. When the dark matter channel is not kinematically allowed the electron and muon searches can cover the mass range of $M_{Z'}=(400-2000)$ GeV. One then can find different most sensitive channels for the rest of the parameter space. However, the tiny values of the couplings and the large mass of the $Z'$ make the total cross section small enough so it lies under the reach of the experimental searches.

The third and last scenario corresponds to the opposite to Scenario 2. In this case the values of the parameters are chosen in such a way that the vector and axial couplings of the leptons to the $Z'$ vanish. This can be seen in the results of the right plot in figure.~\ref{fig:stu_1}. The most sensitive channels in all the parameter space are the monojet, $b\bar{b}$, $t\bar{t}$ and dijet. The monojet region here is very clear. When the dark matter decay channel is kinematically open the monojet search excludes up to masses of $M_{Z'}$. In this mass range but for a non-kinematically allowed dark matter channel the two most sensitive channels are $t\bar{t}$ and dijet, but they are not powerful enough to exclude. In the range of masses between $M_{Z'}=(1-4.5)$ TeV the $b\bar{b}$ search is the most sensitive allowing us to exclude masses that are lower than $M_{Z'}\lesssim 3.2$ TeV.

%
%
\section{Conclusions}
\label{sec:conclu}
In this work we have presented the expansion of the code $Z'$ explorer to further include missing energy searches, where $Z'$ plays the role of a mediator between the dark and visible sectors. We have implemented within the existing $Z'$-explorer framework, which already included all LHC searches for $Z'$ bosons with masses above the weak scale, the ATLAS mono-jet study with 139 fb$^{-1}$ of data. We have found a good agreement between our simulations and the ATLAS results, which allowed us to make an exploratory journey of several $Z'$ models. 

In passing, we also commented on the difficulties encountered in the reinterpretation of the mono-jet study in view of the available reinterpretation material. While we had indeed found a good agreement, the material at our disposal was not always sufficient. For instance, we have benefited from the fact that two different versions of this study used each a different single benchmark point. Only having had one benchmark, the validation of the study would have lacked robustness. We would have also benefited from having access to the 95 \% C.L. upper limits instead of only having the 95 \% C.L. exclusion contours, and it would have been desirable that both exclusive and inclusive selections results were presented for the upper limits and exclusion contours.

We have studied the non-trivial interplay between the visible and invisible channels for dark matter, and showed how they depend on the different coupling choices. In a first step, we have take a bottom-up approach, considering the different $Z'$ couplings as free, independent  parameters. We have departed from the simplifying assumptions of universal couplings for quarks and leptons, showing the rich palette of phenomenological possibilities. We have also explored the impact of the dark matter coupling neither vectorially nor axially to the dark matter particle, showing that even for a non-trivial configuration that involves more channels $Z'$ arises as a powerful tool to test the different regions of the parameter space. These excursions in the simplified model space highlight the importance of having a broad program of experimental searches, as no single channel dominates over the whole parameter space.

However, in explicit $Z'$ models which derive from top-down considerations, there often exists correlations among the different couplings. We have then considered a Stückelberg portal that are obtained as low energy effective actions in some string compactifications with intersecting branes in type IIa string theory. In this construction the $Z'$ couples to the matter fields via a combination of four different charges. Consequently, the matter fields the coupling structure is more complicated involving more parameters and correlations between couplings.  With the aid of $Z'$-explorer 2.0, we have derived the exclusion limits in the $M_{Z'} - m_{\chi}$ plane for different benchmark scenarios. Even for a more complex coupling construction $Z'$-explorer is to be a powerful tool. While different parameter choices in the Stückelberg portal seem to prefer different visible channels, the mono-jet channeld added in this version appears ubiquitiously in the three scenarios explored. This specific search then has a crucial role in testing models that have $Z'$ when the dark matter channel is open. 

Along the way, we have also highlighted the potential improvements that could be added to the code, for example the addition of flavor bounds, considering $Z'$ in the GeV range (known as \emph{dark photons} in the jargon, they are well covered by DarkCast~\cite{Ilten:2018crw}), extrapolating the results to the High-Luminosity LHC (HL-LHC and future colliders), or enlarging the particle content of the dark sector. We leave this interesting options, among others, as possibles avenues for improvement for a new version.


\section*{Acknowledgements:} 

We would like to thank Ezequiel Alvarez for his collaboration in early stages of this work.  We are indebted to Benjamin Fuks and Dipan Sengupta for correspondence regarding the MadAnalysis5 PAD implementation of reference~\cite{ATLAS:2017bfj} and to Boyu Gao for useful cross-checks of the event generation setup. We would also like to thank Giuliano Gustavino, Valerio Ippolito and Steven Worm for useful communication regarding technical details of the ATLAS analysis, and Marie-H\'el\`ene Genest and Nishita Desai for useful discussions. We also thank Diego Mouriño for assistance with the code. JZ is supported by the {\it Generalitat Valenciana} (Spain) through the {\it plan GenT} program (CIDEGENT/2019/068). VML is funded by the Deutsche
Forschungsgemeinschaft (DFG, German Research Foundation) under Germany's Excellence
Strategy - EXC 2121 ``Quantum Universe'' - 39083330. We also want to thank the organizers of the ``(Re)interpreting the results of new physics searches at the LHC'' and ``Río de la Plata Ph-Exp Institute'' workshops and the IFLP where the starting point of this project took off.

\begin{appendix}
 \section{$Z'$ explorer 2.0: implementation details}
 \label{app:code}
 \subsection{Backend}
\label{bend}

As previously described, Z'-explorer 2.0 extend the scope of the software's first version to include final states with missing transverse energy. The implementation for the SM (\emph{visible}) channels remains unchanged, see~\cite{Alvarez:2020yim} for further details. In this version we have modifieded some of the visible channels to include the most sensitive searches up to date. The updated references for each visible channel are: $jj$ \cite{Aad:2019hjw}, $b\bar{b}$ \cite{Aad:2019hjw}, $t\bar{t}$ \cite{Sirunyan:2018ryr}, $e^{+}e^{-}$ \cite{CMS:2021ctt}, $\mu^{+}\mu^{-}$ \cite{CMS:2021ctt}, $\tau^{+}\tau^{-}$ \cite{Aaboud:2017sjh}, $W^{+}W^{-}$ \cite{CMS:2021klu}, $Zh$ \cite{Sirunyan:2019vgt}. Therefore, in this appendix we focus on the implementation of the new DM final state.

To compute the production cross section in the new decay channel, the software uses previously generated and recorded the leading order (LO) $p p \to Z' j$ cross section (instead of just $q \bar{q} \to Z'$, used for visible channels) at $\sqrt{s}=13$ TeV with {\tt MadGraph5\_aMC@NLO}, customizing the UFO-model described in section~\ref{sec:model}, to set $g_{q_A}=g_{q_V}=1$ for only one quark in the proton each time, for $M_{Z'}\in[0.01,2,5]$ TeV, with a step of 0.01 TeV. We have explicitly verified that the initial states with $b$ quarks have a negligible contribution to this process. We employ default settings of {\tt MadGraph5\_aMC@NLO}, and in particular the fast detector simulation uses the ATLAS default card from {\tt Delphes 3.4.2}. 

These cross sections grids, stored in the repository as {\tt /cards/AXIAL(VEC)\_Zpj}, are invoked during the program execution: the predicted production cross section for a benchmark point is simply the sum of the four contributions of quarks ($u$, $d$, $c$ and $s$), each of them adjusted by the sum of the corresponding squared couplings. That is
\begin{equation}
\sigma(p p \to Z' j)^{g_{q_A},\, g_{q_V}}=\sum\limits_{q}\left[\sigma(p p \to Z' j)^{g_{q_A}=1}(g_{q_A})^2+\sigma(p p \to Z' j)^{g_{q_V}=1}(g_{q_V})^2\right]
\end{equation}
The software selects inside the simulations the record with the mass $M_{Z'}$ that is closest to the one in the input card at the corresponding benchmark point. The axial and vector couplings are directly estimated from the chiral coupling in the input card (``incard'')
\be
g_{f_V}=\frac{1}{2}(g_{f_R}+g_{f_L}) \, , \qquad
g_{f_A}=\frac{1}{2}(g_{f_R}-g_{f_L}) \, .
\ee

The production cross-section times branching ratio ($\mathcal{BR}(Z' \to \chi \bar{\chi})$) estimated for the given benchmark point are used to re-scale the experimental limits in~\cite{ATLAS:2021kxv} for the axial-vector mediator scenario, within the NWA.\footnote{The case for the vector mediator can be trivially derived from the axial one, so we will describe the latter only, for simplicity.} The number of events in each exclusive signal region (coming from our recast previously described) are stored for the set of points in the ($M_{Z'}, \, m_{\chi}$) plane mentioned in section~\ref{sec:ATLAS}, in different text files in {\tt /cards/DM/AXIAL} folder. The software selects the file for the ($M_{Z'}, \, m_{\chi}$) pair that is closest to the one in the input card at the corresponding benchmark point, through the determination of the euclidean norm. As mentioned in the main text, this is the main difference with respect to the first version: having upper limits and exclusion countours in 2 dimensions instead of only one. With the corresponding events selected and properly re-scaled, the software selects the most sensitive EM signal region comparing to background (also stored in {\tt /cards/DM/}), and estimates the strength for the $\chi \bar{\chi}$ channel.  


\subsection{Frontend}
\label{fend}
The input card ({\tt /incard/card\_1.dat}) required by in Z'-explorer 2.0 is quite similar to the one needed in its first version. The user must provide the old requirements for the visible channels: $M_{Z'}$ (in TeV), the $Z'$ couplings to all SM-fermions, except for neutrinos, which information is required through the partial width $\Gamma_{\nu\nu}$, as for the $W^{+}W^{-}$ ($\Gamma_{WW}$) and the $Zh$ ($\Gamma_{Zh}$) decay channels. These entries are required as {\it partial width} since there is not a unique Lorentz structure in their couplings. Additionally, for the new channel, we require the fermionic DM mass $m_\chi$ (in TeV), and its couplings to $Z'$. The total width to other non-SM particles can be added in the computation as $\Gamma_{xx}$. Each entry is one row (with twenty-six columns), and corresponds to one benchmark point within the model to be tested by $Z'$-explorer 2.0. The NP parameters should be specified in the following order (with each column separated by spaces)
{\small
\begin{equation}
\hspace{-0.25cm} M_{Z'} \hspace{0.1cm} g_{u_L} \hspace{0.1cm} g_{u_R} \hspace{0.1cm} g_{d_L} \hspace{0.1cm} g_{d_R}\hspace{0.1cm} g_{c_L} \hspace{0.1cm} g_{c_R} \hspace{0.1cm} g_{s_L} \hspace{0.1cm} g_{s_R}\hspace{0.1cm} g_{b_L} \hspace{0.1cm} g_{b_R} \hspace{0.1cm} g_{t_L} \hspace{0.1cm} g_{t_R}
\hspace{0.1cm} g_{e_L} \hspace{0.1cm} g_{e_R} \hspace{0.1cm} g_{\mu_L} \hspace{0.1cm} g_{\mu_R}\hspace{0.1cm} g_{\tau_L}\hspace{0.1cm}g_{\tau_R}\hspace{0.1cm}\Gamma_{\nu\nu}\hspace{0.1cm}\Gamma_{WW}\hspace{0.1cm}\Gamma_{Zh}\hspace{0.1cm}m_{\chi}\hspace{0.1cm}g_{\chi_L}\hspace{0.1cm}g_{\chi_R}\hspace{0.1cm}\Gamma_{xx}
\nonumber
\end{equation}
}
where $g_{f_L}$ ($g_{f_R}$) is the coupling of $Z'$ to the corresponding Left (Right) fermion. To obtain the right form of the $g_{f_{L/R}}$ couplings from a given NP model, interactions between $Z'$ and SM fermions must be written in the parametrization of equation~\eqref{eq:Zplag}. 

The instruction for running the software remains unchanged. In the command line type
\begin{verbatim}
>./program.out
\end{verbatim}

The software displays on command shell a summary of the tasks performed during execution. The addition of the new invisible channel does not significantly change the speed of the performance from the first version ($\mathcal{O}(10^3)$ points can be processed in no more than 2 seconds in a regular CPU). Z'-explorer 2.0 prints in the output file ({\tt /output/1.dat}) the incard information for each benchmark point (to ease the after-processing), followed by  
{\small
\begin{equation}
\mathcal{S}_{jj}\hspace{0.2cm}\mathcal{S}_{bb}\hspace{0.2cm}\mathcal{S}_{tt}\hspace{0.2cm}\mathcal{S}_{ee}\hspace{0.2cm}\mathcal{S}_{\mu\mu}\hspace{0.2cm}\mathcal{S}_{\tau\tau}\hspace{0.2cm}\mathcal{S}_{\nu\nu}\hspace{0.2cm}\mathcal{S}_{WW}\hspace{0.2cm}\mathcal{S}_{Zh}\hspace{0.2cm}\mathcal{S}_{\chi\chi}\hspace{0.2cm}\mathcal{S}_{xx}\hspace{0.2cm}\Gamma_{Z'}\hspace{0.2cm}WARNING:\Gamma_{Z'}>5%
\nonumber
\end{equation}
}where $\mathcal{S}$ is the calculated strength in each channel (already defined in~\cite{Alvarez:2020yim}), $\Gamma_{Z'}$ is the $Z'$ total width, and the last column is a warning for when the input parameters invalidate the narrow width approximation (displays $1$ if $\Gamma_{Z'}>5\%$ and $0$ otherwise). Nonetheless, this warning does not halt the execution, so the resulting rate is computed using NWA even if it is not a good approximation. The parameter $\mathcal{S}_{xx}$ can be regarded as a dummy variable to eventually add experimental information of other possible non-SM $Z'$ channels. For further details about running and the additional information that can be extracted from Z'-explorer 2.0, visit the GitHub repository~\cite{github}.

\end{appendix}

\bibliographystyle{JHEP}
\bibliography{Zexpl2}

\end{document}